\documentclass[onecolumn, draftcls, 12pt]{IEEEtran}

\ifCLASSINFOpdf
  \usepackage[pdftex]{graphicx}
  \DeclareGraphicsExtensions{.pdf,.jpeg,.png}
\else
  \usepackage[dvips]{graphicx}
\fi
\usepackage{bm}
\usepackage{threeparttable}
\usepackage{epstopdf}
\usepackage[compress,nospace]{cite}
\usepackage{amsthm}
\usepackage{stfloats}
\usepackage[cmex10]{amsmath}
\usepackage{amsmath}
\usepackage{cases}
\usepackage{siunitx}
\usepackage{amssymb}
\usepackage{wasysym}
\usepackage{algorithm}
\usepackage{algorithmic}
\usepackage[colorlinks,linkcolor=blue,anchorcolor=blue,citecolor=blue,bookmarks=blue]{hyperref}
\usepackage{array}
\usepackage{url}
\usepackage{color}
\usepackage{subfigure}
\usepackage{verbatim}
\usepackage{multicol}  
\usepackage{multirow} 
\usepackage{booktabs}
\usepackage{makecell}
\allowdisplaybreaks[4]
\usepackage{setspace}
\begin{document}

\title{Learnable Model-Driven Performance Prediction and Optimization for Imperfect MIMO System: Framework and Application}

\author{Fan~Meng, Shengheng~Liu,~\IEEEmembership{Senior Member,~IEEE}, \\Yongming~Huang,~\IEEEmembership{Senior Member,~IEEE}, Zhaohua~Lu
	\thanks{This work was supported in part by the National Key R\&D Program of China under Grant 2020YFB1806600 and the National Natural Science Foundation of China under Grant No. 62001103. Part of this work has been submitted for possible presentation at the IEEE Global Communications Conference (GLOBECOM): Mobile and Wireless Networks Symposium, Rio de Janeiro, Brazil, December 2022 \cite{GlobeCom22}. (Corresponding authors: Y.\ Huang and S.\ Liu)}
	\thanks{F. Meng, S. Liu and Y. Huang are with the Purple Mountain Laboratories, Nanjing 211111, China (e-mail: mengfan@pmlabs.com.cn; s.liu@seu.edu.cn; huangym@seu.edu.cn). S. Liu and Y. Huang are also with the School of Information Science and Engineering, Southeast University, Nanjing 210096, China. Z. Lu is with the ZTE Corporation and State Key Laboratory of Mobile Network and Mobile Multimedia Technology, Shenzhen 518057, China (e-mail: lu.zhaohua@zte.com.cn).}
}

\markboth{Submitted to IEEE Transaction on Wireless Communications}
{Shell \MakeLowercase{\textit{et al.}}: Bare Demo of IEEEtran.cls for Journals}

\maketitle

\begin{abstract}
	
State-of-the-art schemes for performance analysis and optimization of multiple-input multiple-output systems generally experience degradation or even become invalid in dynamic complex scenarios with unknown interference and channel state information (CSI) uncertainty. To adapt to the challenging settings and better accomplish these network auto-tuning tasks, we propose a generic learnable model-driven framework in this paper. To explain how the proposed framework works, we consider regularized zero-forcing precoding as a usage instance and design a light-weight neural network for refined prediction of sum rate and detection error based on coarse model-driven approximations. Then, we estimate the CSI uncertainty on the learned predictor in an iterative manner and, on this basis, optimize the transmit regularization term and subsequent receive power scaling factors. A deep unfolded projected gradient descent based algorithm is proposed for power scaling, which achieves favorable trade-off between convergence rate and robustness. 

\end{abstract}


\section{Introduction}\label{sec:introduction}

Deep learning (DL) is regarded as an enabling technology for future wireless mobile network, and has gained extensive attention~\cite{8723067, 8807386, 9210812, 9257198}. The learning-based approaches are data-driven and inherently applicable for the scenarios with imperfect models and/or intractable problems, meanwhile the model-driven methods usually fail for the complex changing environment~\cite{8054694, 9120241, 9650869}. DL uses deep neural networks (NNs) as function approximators, with a particular emphasis on stacking many layers of structurally similar components. Parametric models with deep structures have formidable nonlinear mapping capability to learn extremely complex functions, but at the expense of overwhelming model complexity, high dependence on data, lack of interpretability and performance guarantee. High complexity results in obstacles for practical implementation in wireless communications, and thus developing low-complexity NNs with very small trainable parameter and training data sets is significant. Therefore, embedding learnable modules into the existing model-based system or designing a specific NN by domain knowledge in communications, can combine the advantages of both paradigms and achieve better performance~\cite{8715338, 8742579}. In physical (PHY) layer, lots of studies have investigated further optimizations of model-based algorithms with data-driven methods, in many aspects such as precoding~\cite{9115718, 9125984, 9210124, 9246287}, beam tracking~\cite{9069211, 9269463}, beam prediction~\cite{9598911}, channel estimation~\cite{9186336}, and signal detection~\cite{9018199}.

\subsection{Related Work}

Learning-enabled performance prediction (PP) and optimization are studied in media access control (MAC) and higher network layer, due to the complexity of explicit modeling~\cite{8253758}. Shi et al.~\cite{9219163} propose to directly compute ergodic sum rate (SR) from statistical channel state information (CSI) via a hybrid deep NN. Then, the predicted SR is utilized for multi-cell user scheduling. Simulations show that the computational burden is reduced while maintaining nearly the same performance as that of the deterministic equivalent based method. The future long-term PP, i.e., the cumulative sum of time-varying data rates, is considered in~\cite{8941039}. A proactive deep reinforcement learning method is proposed, wherein handover timings are optimized while obstacle-caused data rate degradations are predicted before the degradations occur. To associate key performance indicators and key quality indicators in cellular network, a learning-enabled quantitative association rule mining method composed of sliding window partitioning and random forest is developed~\cite{8903320}. Besides, the PP problems in wireless mobile networks including outage probability prediction for network performance evaluation~\cite{8877818}, traffic prediction and random access control optimization~\cite{9422330}, are addressed in DL approaches. 

In PHY, the DL methods can be further enhanced with well-studied model-based results. Thrane et al.~\cite{8950164} investigate channel modeling for system coverage. Based on a simple path loss model, the DL techniques with satellite images are further used to realize accurate path loss prediction. Additionally, the model-based performance analysis for multiple-input multiple-output (MIMO) system has been widely investigated~\cite{6172680, 6415386, 6415388, 6457363}, and the theoretical results are then utilized for numerous optimizations including antenna number, regularization parameter, power allocation, and user number. To the best of our knowledge, inter-discipline of PHY performance analysis and machine learning has rarely been addressed in the literature.

Theoretical results derived from performance analysis provide significant guidance for the subsequent optimization design, but the effectiveness can be significantly degraded by imperfect models. For example, finite quantization A/D are widely deployed in communication systems, but incorporating hardware imperfections in the analysis is difficult~\cite{4537203}; Cramer-Rao lower bound (CRLB) is only achievable for unbiased estimation in high signal-to-noise ratio (SNR) range, but CRLB is not tight in low SNR range~\cite{1611097}; the deterministic equivalence of system capacity derived by random matrix theory holds true for sufficient large antennas, but the approximation with limited antennas can be loose~\cite{7479496}.

\subsection{Motivation and Contribution}

In this work, we first propose a general model-based learnable framework for PP and subsequent optimization in imperfect MIMO system. In principle, the learned PP model is a digital twin~\cite{9374645} of the true scenario, and the optimizations are carried out in this virtual environment. Specifically, using the coarse results of theoretical derivation, the mapping from dependent variables to predict performance is then refined by a light-weight NN. With utilization of domain knowledge, the derivation performs as a fixed operator. To make up the inaccurate results, the learnable NN is trained to approximate the real performance using backward propagation. In the cascaded optimizations stage, utility functions with respect to optimization variables are optimized on the learned PP model.

Secondly, we investigate an instance under the proposed framework. Concretely, in an imperfect multi-user multiple-input single-output (MU-MISO) system with dynamic broadcast channel and unknown CSI uncertainty, we consider expected transmit SR maximization and receive signal detection mean square error (MSE) minimization on a learned PP model. Against channel uncertainty, the SR is maximized by tuning the regularization term $ \alpha $ in regularized zero-forcing (RZF) precoding, and the subsequent detection MSE is minimized by receive power scaling. In engineering, the RZF precoding can be optimized without knowledge of CSI uncertainty by conventional line search of $ \alpha $. However, the searching requires repeated interactions with the environment, which is challenging for dynamic scenarios. The SR maximization is investigated in \cite{6172680}, an approximated optimal closed-form RZF design is derived, but it has following shortcomings. (i) Unknown CSI uncertainty: in a frequency division duplexing system, the CSI uncertainty is difficult to be directly obtained since it can be arisen by channel estimation error, pilot contamination, channel quantization, delayed feedback, channel fast fading, etc. (ii) Intractable problems: the closed-form optimum only holds with the assumption of infinite and uncorrelated antennas, equal CSI uncertainty for all users. (iii) Imperfect models: the practical scenario can be imperfect for the established system model. Therefore, we need a fast CSI uncertainty sensing and estimation method for unaware CSI, and data-driven methods for intractable problems and imperfect models. Inspired by the works in \cite{6172680}, our main contributions are summarized as follows.
\begin{itemize}
\item We propose a general learnable model-based framework for PP and subsequent optimization in imperfect MIMO system. Under the PP framework, we design a low-complexity predictor of SR and detection MSE. Besides, we derive the deterministic equivalent MSE of receive signal detection.
\item We propose a procedure for transmit CSI uncertainty sensing, with very low interactive information. Using the learned predictor, we prove that the CSI uncertainty problem is non-convex, then use iterative searching method for CSI uncertainty estimation.
\item We use the learned predictor and estimated CSI uncertainty for optimizations. For transmit SR maximization, we optimize the regularization term in RZF precoding by iterative searching; for receive signal detection MSE minimization, we develop a deep unfolded projected gradient descending (PGD) algorithm for power scaling, to achieve a good trade-off between convergence speed and robustness.
\end{itemize}

The rest of this paper is organized as follows. The system model and the problem formulation are described in Section.~\ref{sec:system}. The framework of learnable model-driven performance prediction is clarified in Section.~\ref{sec:framework}. Furthermore, a learnable RZF beamforming is given in Section.~\ref{sec:example}, including performance indicator sensing, CSI uncertainty estimation, and optimizations. The numerical results are shown in Section.~\ref{sec:simulation}, and the conclusions are drawn in Section.~\ref{sec:conclusion}.

\textit{Notations}: We use lowercase (uppercase) boldface $ \boldsymbol{a}(\boldsymbol{A}) $ to denote a vector (matrix), and $ a $ is a scalar. Calligraphy letter $ \mathcal{A} $ represents a set or some probability distribution. Superscripts $ (\cdot)^{\textup{T}} $, $ (\cdot)^* $ and $ (\cdot)^{\textup{H}} $ represent the transpose, the complex conjugate and the Hermitian transpose, respectively. $ \textup{tr}(\cdot) $ denotes the trace. $ \overline{(\cdot)} $ denotes the cardinality. $ \mathbb{E}\{\cdot\} $ denotes the expectation operator. $ \boldsymbol{I}_N $ denotes the $ N\times N $ identity matrix, and $ \boldsymbol{n} \sim \mathcal{CN}(\boldsymbol{0}, \boldsymbol{I}_N) $ means $ \boldsymbol{n} $ is complex circularly-symmetric Gaussian distributed with zero mean and covariance $ \boldsymbol{I}_N $. $ |\cdot| $ is the absolute operator, $ \|\cdot\|_p $ denotes the $ \ell_p $ norm. $ \odot $ is dot product. $ \mathbb{R} $ and $ \mathbb{C} $ represent the real field and complex field, respectively.

\section{System Model}\label{sec:system}

Consider a downlink MU-MISO broadcast channel where a base station (BS) equipped with $ M $ antennas transmits narrow-band signal to $ K $ single-antenna users. The received signal of $ K $ users is expressed as
\begin{equation}\label{equ:y_k}
\boldsymbol{y} = \boldsymbol{H} \boldsymbol{x} + \boldsymbol{n}
\end{equation}
where element $ y_k $ of $ \boldsymbol{y} = [y_1, \cdots, y_K]^{\textup{T}} $ denotes received signal of user $ k $, column vector $ \boldsymbol{h}_k \in \mathbb{C}^{M} $ of channel matrix $ \boldsymbol{H} = [\boldsymbol{h}_1, \cdots, \boldsymbol{h}_K]^{\textup{H}} \in \mathbb{C}^{K \times M} $ denotes channel of user $ k $, $ \boldsymbol{x} \in \mathbb{C}^{M} $ is transmit vector, and $ \boldsymbol{n} $ is complex additional white Gaussian noise (AWGN) following $ \mathcal{CN}(\boldsymbol{0}, \boldsymbol{I}_K) $. Each user channel is correlated, i.e., $ \mathbb{E}[\boldsymbol{h}_k \boldsymbol{h}_k^{\textup{H}}] = \boldsymbol{\Theta}_k $, and $ \boldsymbol{h}_k $ can be expressed as
\begin{equation}\label{equ:h_k}
\boldsymbol{h}_k = \boldsymbol{\Theta}_k^{\frac{1}{2}} \boldsymbol{z}_k
\end{equation}
where $ \boldsymbol{\Theta}_k $ is the channel correlation matrix of user $ k $, which is assumed to be slowly varying compared to the channel coherence time and thus are supposed to be perfectly known to the transmitter; entries in $ \boldsymbol{z}_k $ are independent and identical distributed (i.i.d.), and follow $ \mathcal{CN}(0, 1) $. Moreover, only an imperfect estimate $ \hat{\boldsymbol{h}}_k $ of the true channel $ \boldsymbol{h}_k $ is available at the BS, which is modeled as follows
\begin{equation}\label{equ:h_hat_k}
\hat{\boldsymbol{h}}_k = \boldsymbol{\Theta}_k^{\frac{1}{2}} \Big(\sqrt{1 - \tau_k^2} \boldsymbol{z}_k + \tau_k \boldsymbol{q}_k \Big)
\end{equation}
where $ \boldsymbol{q}_k $ have i.i.d. entries following $ \mathcal{CN}(0, 1) $, and $ \tau_k \in [0, 1] $ reflects the amount of CSI uncertainty.

The transmit vector $ \boldsymbol{x} $ is a linear transformation of user symbol vector $ \boldsymbol{s} = [s_1, \cdots, s_K]^{\textup{T}} $ where $ s_k \sim \mathcal{CN}(0, 1) $, and $ \mathbb{E}\{\textup{tr}(\boldsymbol{s} \boldsymbol{s}^{\textup{H}})\} = \boldsymbol{I}_K $. Specifically, $ \boldsymbol{x} $ can be written as
\begin{equation}\label{equ:x}
\boldsymbol{x} = \boldsymbol{G} \boldsymbol{P}^{\frac{1}{2}} \boldsymbol{s}
\end{equation}
where $ \boldsymbol{P} = \textup{diag}(p_1, \cdots, p_K) $ and $ \boldsymbol{G} = [\boldsymbol{g}_1, \cdots, \boldsymbol{g}_K] \in \mathbb{C}^{M \times K} $ respectively are signal power matrix and beamforming matrix, and the elements $ p_k $ and $ \boldsymbol{g}_k $ respectively are signal power and beamforming vector of user $ k $. The transmit signal is normalized to satisfy the average total power constraint
\begin{equation}\label{equ:P_constraint}
\mathbb{E}\{\|x\|^2\} = \textup{tr}(\boldsymbol{G} \boldsymbol{P} \boldsymbol{G}^{\textup{H}}) \leq P
\end{equation}
where $ P $ denotes the maximal total transmit power.

The regularized zero forcing precoder $ \boldsymbol{G} $ is given by
\begin{equation}\label{equ:G}
\boldsymbol{G} = \xi \underbrace{\big(\hat{\boldsymbol{H}}^{\textup{H}} \hat{\boldsymbol{H}} + M \alpha \boldsymbol{I}_M\big)^{-1}}_{\hat{\boldsymbol{W}}} \hat{\boldsymbol{H}}^{\textup{H}}
\end{equation}
where $ \hat{\boldsymbol{H}} = [\hat{\boldsymbol{h}}_1, \cdots, \hat{\boldsymbol{h}}_K]^{\textup{H}} $, $ \xi $ is a normalization scalar to fulfill the power constraint in \eqref{equ:P_constraint}, and $ \alpha > 0 $ is a regularization term. According to \eqref{equ:P_constraint}, we obtain $ \xi^2 $ as follows
\begin{equation}\label{equ:xi}
\xi^2 = \frac{P}{\textup{tr}\Big(\boldsymbol{P} \hat{\boldsymbol{H}} \big(\hat{\boldsymbol{H}}^{\textup{H}} \hat{\boldsymbol{H}} + M \alpha \boldsymbol{I}_M\big)^{-2} \hat{\boldsymbol{H}}^{\textup{H}}\Big)} \triangleq \frac{P}{\Psi}.
\end{equation}
We denote $ \rho \triangleq \frac{P}{\sigma^2} $ as the SNR. At receiver $ k $, the signal-to-interference plus noise ratio (SINR) $ \gamma_k $ under RZF beamforming and single-user decoding with imperfect CSI, takes the form
\begin{equation}\label{equ:gamma_k}
\gamma_k \triangleq \frac{p_k \big|\boldsymbol{h}_k^{\textup{H}} \hat{\boldsymbol{W}} \hat{\boldsymbol{h}}_k\big|^2}{\boldsymbol{h}_k^{\textup{H}} \hat{\boldsymbol{W}} \hat{\boldsymbol{H}}_{[k]}^{\textup{H}} \boldsymbol{P}_{[k]} \hat{\boldsymbol{H}}_{[k]} \hat{\boldsymbol{W}} \boldsymbol{h}_k + \frac{\Psi}{\rho}}
\end{equation}
where $ \hat{\boldsymbol{H}}_{[k]} = [\hat{\boldsymbol{h}}_1, \cdots, \hat{\boldsymbol{h}}_{k-1}, \hat{\boldsymbol{h}}_{k+1}, \cdots, \hat{\boldsymbol{h}}_K]^{\textup{H}} $ and $ \boldsymbol{P}_{[k]} = \textup{diag}(p_1, \cdots, p_{k-1}, p_{k+1}, \cdots, p_K) $. The ergodic SR is defined as
\begin{equation}\label{equ:sumrate}
R \triangleq \mathbb{E}_{\hat{\boldsymbol{H}}} \bigg\{\sum_{k=1}^{K}\log (1 + \gamma_k) \bigg\}.
\end{equation}
For receive signal detection, the user $ k $ should rescale the received signal by a power scaling factor $ u_k $ to obtain an estimation of transmit constellation points. The MSE of user $ k $ is expressed as follows
\begin{align}
\textup{MSE}_k & \triangleq \mathbb{E}_{\boldsymbol{s}, n_k} \bigg\{\Big|u_k y_k - s_k \Big|^2 \bigg\}\nonumber\\
& = \mathbb{E}_{\boldsymbol{s}, n_k} \bigg\{\bigg|u_k \Big(\boldsymbol{h}_k^{\textup{H}} \sum_{j=1}^K \sqrt{p_j} \xi \big(\hat{\boldsymbol{H}}^{\textup{H}} \hat{\boldsymbol{H}} + M \alpha \boldsymbol{I}_M\big) \hat{\boldsymbol{h}}_j^{\textup{H}} s_j + n_k\Big) - s_k\bigg|^2 \bigg\}\nonumber\\
& = \big(u_k \xi \sqrt{p_k} \boldsymbol{h}_k^{\textup{H}} \hat{\boldsymbol{W}} \hat{\boldsymbol{h}}_k - 1\big)^2 + u_k^2 \xi^2 \boldsymbol{h}_k^{\textup{H}} \hat{\boldsymbol{W}} \hat{\boldsymbol{H}}_{[k]}^{\textup{H}} \boldsymbol{P}_{[k]} \hat{\boldsymbol{H}}_{[k]} \hat{\boldsymbol{W}} \boldsymbol{h}_k + u_k^2 \sigma^2\label{equ:MSE_k}
\end{align}
Similarly, we define the ergodic MSE as
\begin{equation}\label{equ:MSE}
\textup{MSE} \triangleq \mathbb{E}_{\hat{\boldsymbol{H}}} \bigg\{\frac{1}{K} \sum_{k=1}^{K} \textup{MSE}_k \bigg\}.
\end{equation}
The SR and MSE are regarded as the performance indicators of the investigated model.

The established model does not match the practical scenario in many ways, and we take two for example. Firstly, it is feasible to obtain a time average of $ \textup{MSE} $ as a statistical average in \eqref{equ:MSE}. Meanwhile, the elements in $ \boldsymbol{s} $ are unknown data symbols, thus the time average is unpresented. As an alternative, a feasible observation of $ \textup{MSE}_{k} $ is given as
\begin{align}\label{equ:mse_r}
\textup{MSE}_{\textup{m}, k} & = \mathbb{E}_{\hat{s}_k, n_k} \bigg\{\Big|u_k y_k - \hat{s}_k \Big|^2 \bigg\}\nonumber\\
& = (1 - P_e) \mathbb{E}_{\hat{s}_k, n_k} \Big\{\big|u_k y_k - \hat{s}_k|_{\hat{s}_k = s_k}\big|^2 \Big\} + P_e \mathbb{E}_{\hat{s}_k, n_k} \Big\{\big|u_k y_k - \hat{s}_k|_{\hat{s}_k \neq s_k}\big|^2 \Big\}\nonumber\\
& = (1 - P_e) \textup{MSE}_k + P_e \mathbb{E}_{\hat{s}_k, n_k} \Big\{\big|u_k y_k - \hat{s}_k|_{\hat{s}_k \neq s_k}\big|^2 \Big\} \leq \textup{MSE}_k
\end{align}
where $ P_e $ is the symbol error rate (SER), $ \hat{s}_k $ is the demodulated symbol of user $ k $. The decision of $ \hat{s}_k $ is the one with the shortest path from the received signal to all constellation points, thus $ \big|u_k y_k - \hat{s}_k|_{\hat{s}_k \neq s_k}\big| \leq \big|u_k y_k - \hat{s}_k|_{\hat{s}_k = s_k}\big| $. When $ \rho \to +\infty $, $ P_e \to 0 $, then $ \textup{MSE}_{\textup{m}, k} \to \textup{MSE}_k $. The time averages of $ \textup{MSE}_{\textup{m}, k} $ are measured at the users and then feedback to the BS.

Secondly, in the complex electromagnetic interference environment, the achievable rate of a link can be significantly reduced by the existing unknown interference. The SNR $ \rho $ is an ideal value, meanwhile the practical $ \rho_{\textup{m}} $ can be ruined by the unknown interference and $ \rho_{\textup{m}} < \rho $.

\section{Learnable Model-driven Framework for Performance Prediction and Optimization}\label{sec:framework}

Without loss of generality, the PP is defined as a map $ f $ from a set of given input $ \boldsymbol{\mathcal{X}} = \{\mathcal{X}_i\}_{i=1}^{\overline{\boldsymbol{\mathcal{X}}}} $ to a set of performance indicators $ \boldsymbol{\mathcal{Y}} = \{\mathcal{Y}_i\}_{i=1}^{\overline{\boldsymbol{\mathcal{Y}}}} $. With learnable parameter set $ \boldsymbol{\Theta}_f $, the learning-enabled function $ f $ is formulated as follows
\begin{equation}\label{equ:g_p}
\boldsymbol{\mathcal{Y}} = f(\boldsymbol{\mathcal{X}}; \boldsymbol{\Theta}_f)
\end{equation}

\subsection{Equivalent Problem Transformation}

Meanwhile, function $ f $ can be too complex to approximate by an end-to-end data-driven method, and thus we equivalently transform the primary map \eqref{equ:g_p} into the following formulation
\begin{subequations}
\begin{align}
\boldsymbol{\mathcal{Y}} & = g(\boldsymbol{\mathcal{X}}; \boldsymbol{\Theta}_g, \boldsymbol{\Theta}_h)\label{equ:g_1}\\
& = \boldsymbol{W}(\boldsymbol{\mathcal{X}}; \boldsymbol{\Theta}_g) h(\boldsymbol{\mathcal{X}}; \boldsymbol{\Theta}_h) + \boldsymbol{b}(\boldsymbol{\mathcal{X}}; \boldsymbol{\Theta}_g)\label{equ:g_2}
\end{align}
\end{subequations}
where function $ h $ is a model-driven formulation and $ \boldsymbol{\Theta}_h $ is a numerical parameter set, function $ g $ is a data-driven map and $ \boldsymbol{\Theta}_g $ is the corresponding learnable parameter set, $ \boldsymbol{W} $ is a weight matrix and $ \boldsymbol{b} $ is a bias vector. When $ h $ has a closed-form, then $ \boldsymbol{\Theta}_h $ is empty, otherwise $ \boldsymbol{\Theta}_h $ is derived by numerical calculation to ensure that $ h $ is closed-formed. $ \boldsymbol{\mathcal{X}} $ is assumed to be sufficient statistics of $ \boldsymbol{\Theta}_h $. Expression \eqref{equ:g_1} is a generalized formulation of the learnable model-driven PP, and \eqref{equ:g_2} is a linear realization of \eqref{equ:g_1}. Furthermore, $ \boldsymbol{\Theta}_h $ can be left out when a closed-form $ h $ is presented, then \eqref{equ:g_2} can be rewritten as
\begin{equation}\label{equ:g_3}
\boldsymbol{\mathcal{Y}} = \boldsymbol{W}(\boldsymbol{\mathcal{X}}; \boldsymbol{\Theta}_g) h(\boldsymbol{\mathcal{X}}) + \boldsymbol{b}(\boldsymbol{\mathcal{X}}; \boldsymbol{\Theta}_g)
\end{equation}
Generally the function $ g $ includes a NN-based part and a model-driven part $ h $ which is not learnable. When entries in $ \boldsymbol{\mathcal{Y}} $ are uncorrelated, then $ \boldsymbol{W} $ is reduced to be a vector $ \boldsymbol{w} $. Then, the linear realization can be rewritten as\footnote{The possible $ \boldsymbol{\Theta}_h $ is omitted for a simplified representation.}
\begin{equation}\label{equ:g_4}
\boldsymbol{\mathcal{Y}} = \boldsymbol{w}(\boldsymbol{\mathcal{X}}; \boldsymbol{\Theta}_g) \odot h(\boldsymbol{\mathcal{X}}) + \boldsymbol{b}(\boldsymbol{\mathcal{X}}; \boldsymbol{\Theta}_g)
\end{equation}
For simplicity, the $ \boldsymbol{W} $ can be further reduced to be an identity matrix which is not learnable, and \eqref{equ:g_3} is represented as
\begin{equation}\label{equ:g_5}
\boldsymbol{\mathcal{Y}} = h(\boldsymbol{\mathcal{X}}) + \boldsymbol{b}(\boldsymbol{\mathcal{X}}; \boldsymbol{\Theta}_g)
\end{equation}
In principle, the proposed NN-enabled $ g $ only learns the difference between the practical map $ f $ and the model-driven PP $ h $. Formulation \eqref{equ:g_5} is similar to residual learning~\cite{Principles}, which was proposed to resolve the training loss degradation problem in very deep networks, by introducing an identity map. Since $ g $ only learns the difference instead of the primary complex map, we regard that manifold of $ g $ is greatly smoother than that of $ f $. To achieve the same upper bound of training loss, $ g $ requires less learnable parameters than $ f $. Thus, computation and storage complexities are reduced, and less training data is required.

Given the labeled training set $ \{\boldsymbol{\mathcal{X}}, \boldsymbol{\mathcal{Y}}_{\textup{m}}\} $, the supervised training problem of learnable model-driven PP is established as a regression as follows
\begin{equation}\label{equ:train_PP}
\begin{split}
& \min_{\boldsymbol{\Theta}_g} \mathbb{E}_{\boldsymbol{\mathcal{Y}}_{\textup{m}} \sim \boldsymbol{\mathcal{Y}}} \Big\{\big\| \boldsymbol{\mathcal{Y}}_{\textup{p}} - \boldsymbol{\mathcal{Y}}_{\textup{m}} \big\|_2^2\Big\}\\
& \, \textup{s.t.} \,\,\,\, \boldsymbol{\mathcal{Y}}_{\textup{p}} = g(\boldsymbol{\mathcal{X}}; \boldsymbol{\Theta}_g)
\end{split}
\end{equation}
where the subscripts $ (\cdot)_{\textup{m}} $ and $ (\cdot)_{\textup{p}} $ respectively identify the measured and predicted indicators.

\subsection{Performance Predictor Analysis}

Theoretical analysis on the learned PP is difficult and unclear, since that part of $ g $ is NN-enabled and properties of $ g $ heavily rely on the numerical values in $ \boldsymbol{\Theta}_g $. For example, the linear transformed \eqref{equ:g_2} can have different function properties compared to model-driven PP $ h $. Meanwhile, the proposed learnable PP is only targeted for the gap caused by imperfect models and/or intractable problems, and can hardly change the trend of the primary PPs. To carry out a feasible analysis on PPs, we propose an assumption as follows
\begin{itemize}
\item[] \textbf{Assumption:} \textit{Learnable model-driven PP and primary model-driven PP are homogeneous.}
\end{itemize}
More concretely, this conjecture indicates that learning-aided PP has the same properties as that of the model-driven PP, e.g., concavity and convexity, monotonicity, number of stationary points, etc. Meanwhile, the exact solutions of these points can be different. Under this conjecture, the optimizations with respect to the input subset can then be realized with the trained predictor, by PGD or searching methods. The learned PP is profound for numerous applications, and we will give an concrete application in the next section.

\section{Application: Intelligent Transmit Beamforming and Receive Detection}\label{sec:example}

\begin{figure}[h!]
	\centering
	\includegraphics[width = 4.5in]{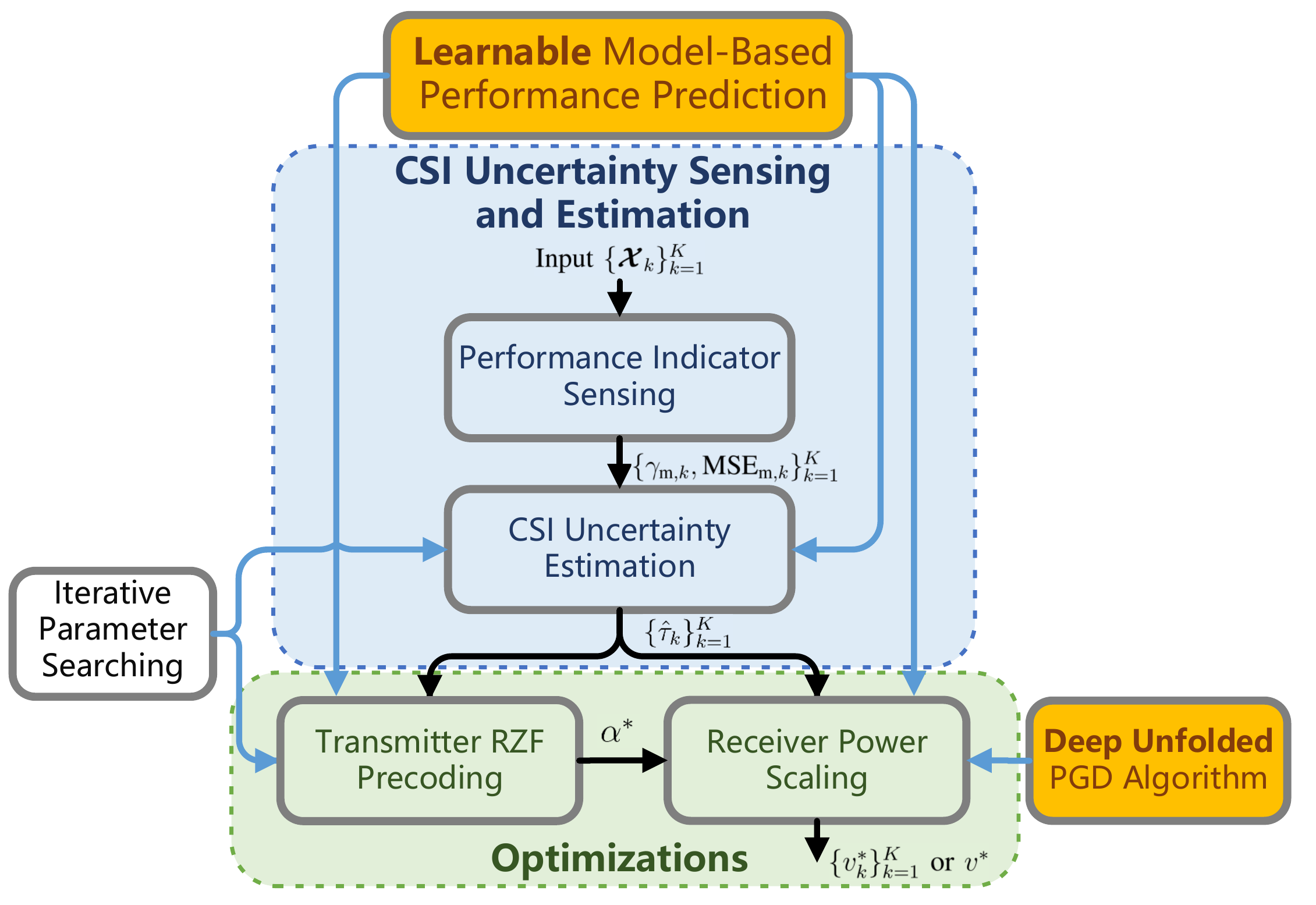}
	\caption{The illustrative procedure of the proposed learning-enabled CSI uncertainty unaware transmit RZF beamforming and the cascaded receive power scaling.}
	\label{fig:overall_framework}
\end{figure}

We investigate the learning-enabled CSI uncertainty unaware transmit RZF beamforming and the cascaded receive power scaling. As shown in Fig.~\ref{fig:overall_framework}, the procedure is composed of two stages: CSI uncertainty sensing and estimation including performance indicator sensing and CSI uncertainty estimation, optimizations including transmit RZF precoding and receive power scaling.

\begin{figure}[h!]
	\centering
	\includegraphics[width = 3.6in]{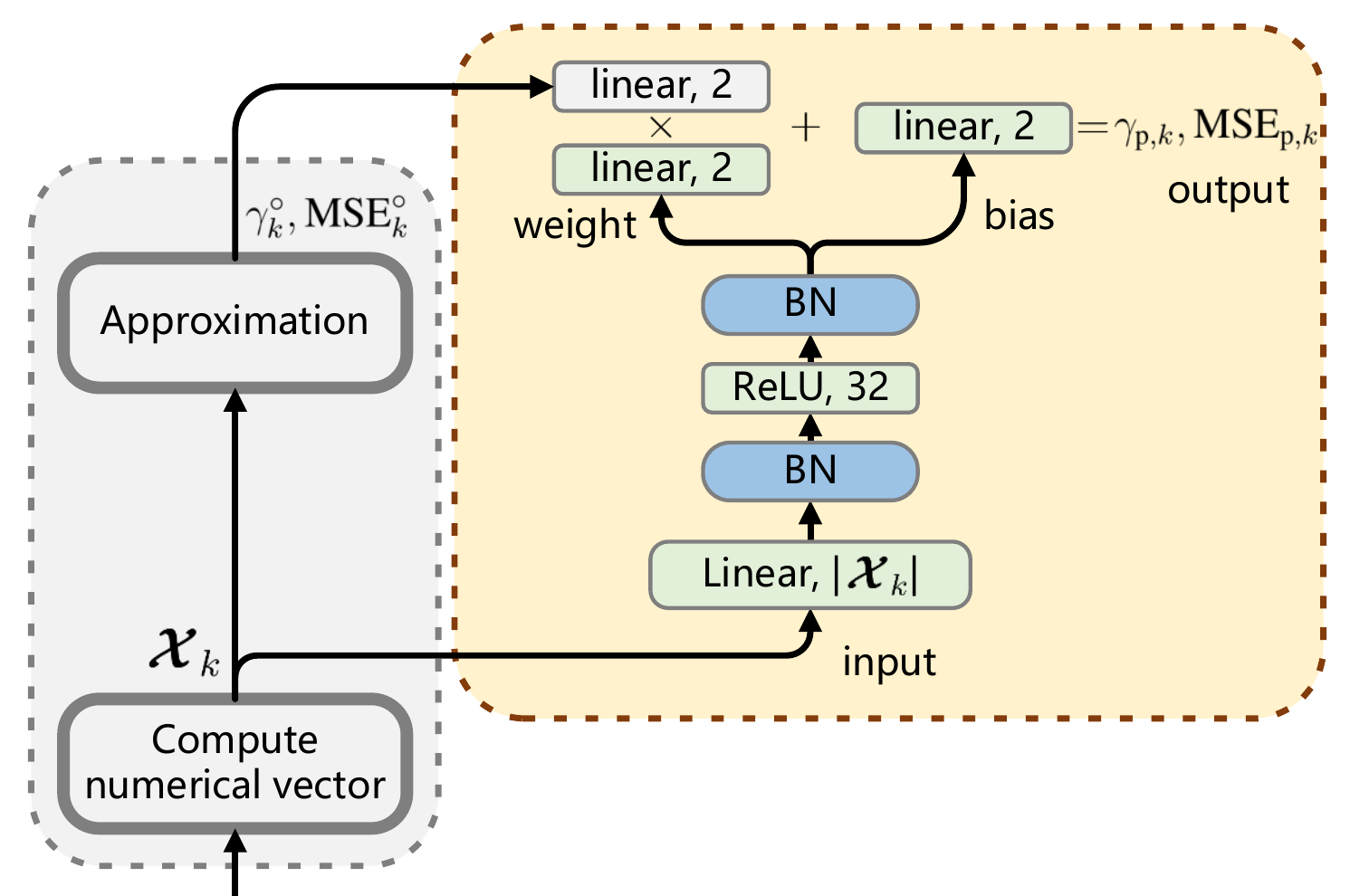}
	\caption{The illustrative topology of learnable model-based performance prediction. The learnable part is labeled with orange.}
	\label{fig:predict_net}
\end{figure}

The core module in Fig.~\ref{fig:overall_framework} is the learnable model-based PP, which provides the PP model for CSI uncertainty estimation and further optimizations. The concrete structure of PP model is depicted in Fig.~\ref{fig:predict_net}, where the model-based part is labeled with gray, and the learnable part is realized by an NN and is labeled with orange. According to Appendix~\ref{sec:appendix_b}, the model-based part computes the deterministic equivalent results and feeds them into the learnable part, and the learnable part generates a final predict by \eqref{equ:g_4} with input $ \boldsymbol{\mathcal{X}}_k $ and the model-based results. Particularly, the notation `BN' denotes batch normalization (BN), notation `ReLU' denotes rectified linear unit (ReLU), and the digit on the right side is the node number.

\subsection{Performance Indicator Sensing}

In this subsection, the BS collects performance indicators from the users to further estimate the CSI uncertainty. According to ergodic theorem, the statistical performance indicator set $ \boldsymbol{\mathcal{Y}} $ is generated at the receivers by time averaging. We make the following assumptions on interval of the time averaging:
\begin{enumerate}
\item Within an interval, statistical characteristics of the environment is unchanged.\label{ass:1}
\item The duration of an interval is sufficient large and feedbacks converge to some constants.\label{ass:2}
\item Across different intervals, statistical characteristics of the environment can be changed.\label{ass:3}
\end{enumerate}
With assumption \ref{ass:1}) and \ref{ass:2}), the feedbacks converge to the corresponding statistical average, i.e., $ \boldsymbol{\mathcal{Y}}_{\textup{m}} \to \boldsymbol{\mathcal{Y}} $.

\begin{figure}[h!]
	\centering
	\includegraphics[width = 3.2in]{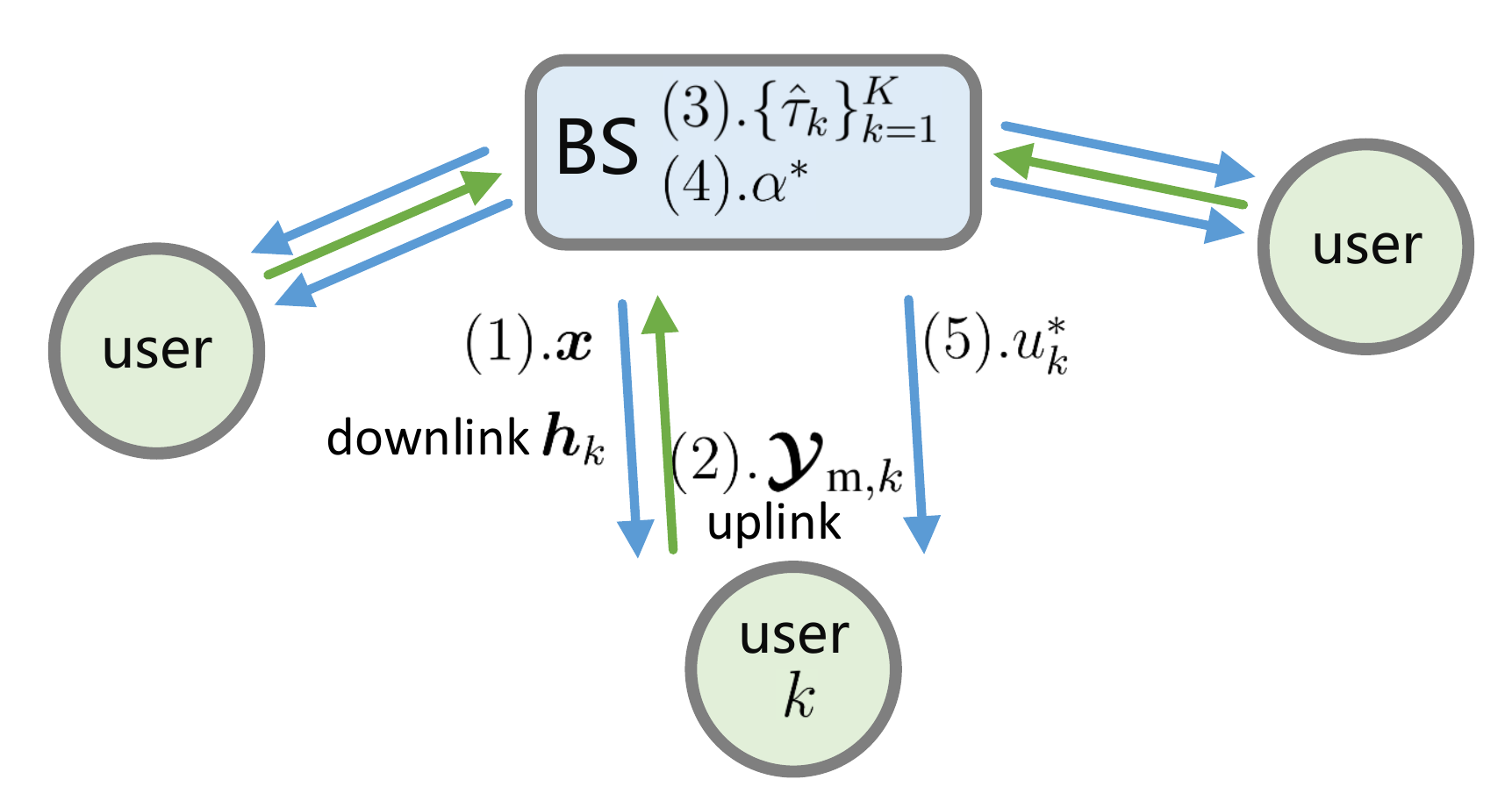}
	\caption{An illustration of information interactions between the BS and users.}
	\label{fig:information_interaction}
\end{figure}

As shown in Fig.~\ref{fig:information_interaction}, the performance indicator sensing is composed of two stages. Firstly,  during the $ t $-th interval, the BS transmits precoded signals with a fixed $ \alpha $, which can be the latest $ \alpha^{t-1} $ or randomly initialized if $ \alpha^{t-1} $ is unpresented. Meanwhile, the users receive signals and compute time average SINRs $ \{\gamma_{\textup{m}, k}^t\}_{k=1}^K $ and MSEs $ \{\textup{MSE}_{\textup{m}, k}^t\}_{k=1}^K $, i.e., $ \boldsymbol{\mathcal{Y}}_{\textup{m}}^t = \{[\gamma_{\textup{m}, k}^t, \textup{MSE}_{\textup{m}, k}^t]^{\textup{T}}\}_{k=1}^K $. Secondly, the users feedback $ \boldsymbol{\mathcal{Y}}_{\textup{m}}^t $ to the central BS via backhaul links.\footnote{In the remainder, we leave out the superscript $ (\cdot)^t $ for convenience.}

\subsection{CSI Uncertainty Estimation}

At the BS side, CSI uncertainty estimation is then carried out with feedback performance indicators in Fig.~\ref{fig:information_interaction}. CSI uncertainties of different users are decoupled, thus the corresponding estimation can be carried out individually. For user $ k $, the estimation problem is formulated as
\begin{equation}
\begin{split}\label{equ:P_tau}
& \min_{\tau_k} \underbrace{\big\| \boldsymbol{\mathcal{Y}}_{\textup{p}, k} - \boldsymbol{\mathcal{Y}}_{\textup{m}, k} \big\|_2^2}_{J_k}\\
& \, \textup{s.t.} \,\,\,\, \boldsymbol{\mathcal{Y}}_{\textup{p}, k} = g(\boldsymbol{\mathcal{X}}_k; \boldsymbol{\Theta}_g)
\end{split}
\end{equation}
where $ \tau_k $ denotes CSI uncertainty, and input $ \boldsymbol{\mathcal{X}}_k $ for different cases is listed in Table~\ref{tab:input}. For simplification, we define that $ \boldsymbol{\mathcal{X}}_k $ is concatenated by three vectors: a prior vector $ \boldsymbol{P}_k $ including variables remain unchanged under varying cases; an optimization vector $ \boldsymbol{O}_k = [\alpha, \tau_k, v_k]^{\textup{T}} $; and a numerical vector $ \boldsymbol{N}_k $ which is determined by the other two vectors, according to the model-based analysis in Appendix~\ref{sec:appendix_b}. From the analysis on $ J_k $ in Appendix~\ref{sec:appendix_c}, \eqref{equ:P_tau} is non-convex over $ \tau_k $ even in the case where $ \boldsymbol{\Theta}_k = \boldsymbol{I}_M $ and $ \tau_k = \tau $. Meanwhile, the closed-form of $ J_k $ is presented, a feasible method is to iteratively search the estimate $ \hat{\tau}_k $ which minimizes $ J_k $ over the learned function $ g $. The detailed CSI uncertainty estimation algorithm is given in Algorithm~\ref{alg:CUS}.

\begin{table}[h!]
	\centering
	\setlength{\tabcolsep}{2.0mm}{
	\small
	\centering
	\caption{Input vectors in four cases} 
	\begin{tabular}{c|ccc} 
		\toprule  
		\multirow{2}{*}{case} & \multicolumn{3}{c}{input $ \boldsymbol{\mathcal{X}}_k $} \\ 
		\cline{2-4}
		& prior vector $ \boldsymbol{P}_k $ & numerical vector $ \boldsymbol{N}_k $ & optimization vector $ \boldsymbol{O}_k $\\
		\midrule
		1. $ \boldsymbol{\Theta}_k, \tau_k $ & $ [M, K, P, p_k, \sigma^2]^{\textup{T}} $ & $ [e_k, \Upsilon_k^{\circ}, \Psi^{\circ}]^{\textup{T}} $ & $ [\alpha, \tau_k, v_k]^{\textup{T}} $\\
		2. $ \boldsymbol{\Theta}_k = \boldsymbol{\Theta}, \tau_k $ & $ [M, K, P, p_k, \sigma^2]^{\textup{T}} $ & $ [e, e_{12}, e_{22}]^{\textup{T}} $ & $ [\alpha, \tau_k, v_k]^{\textup{T}} $\\
		3. $ \boldsymbol{\Theta}_k = \boldsymbol{I}_M, \tau_k $ & $ [M, K, P, p_k, \sigma^2]^{\textup{T}} $ & $ [e]^{\textup{T}} $ &  $ [\alpha, \tau_k, v_k]^{\textup{T}} $\\
		4. $ \boldsymbol{\Theta}_k = \boldsymbol{I}_M, \tau_k = \tau $ & $ [M, K, P, \sigma^2]^{\textup{T}} $ & $ / $ & $ [\alpha, \tau, v]^{\textup{T}} $\\
		\bottomrule 
	\end{tabular} 
	\label{tab:input}}
\end{table}

Finally, we replace the CSI uncertainty in $ \boldsymbol{\mathcal{X}}_k $ with the estimate $ \hat{\tau}_k $, while the other entries remain unchanged. The obtained updated $ \boldsymbol{\mathcal{X}}_k $ is the input vector for the next optimizations. 

\begin{algorithm}
	\caption{CSI uncertainty estimation algorithm.}
	\begin{algorithmic}[1]
		\STATE \emph{Input:} Observed $ \boldsymbol{\mathcal{Y}}_{\textup{m}, k} $, input vector $ \boldsymbol{\mathcal{X}}_k $, maximum and minimum CSI uncertainties $ \tau_{\max}, \tau_{\min} $, division number $ N_{\tau} $, maximum iteration $ L_{\tau} $.
		\STATE \emph{Initialization:} Assign the initial maximum and minimum CSI uncertainties $ \tau_{k, \max}^1 = \tau_{\max} $ and $ \tau_{k, \min}^1 = \tau_{\min}, \forall k $.
		\FOR{$ l = 1 $ to $ L_{\tau} $}
		\STATE Derive $ \{\tau^l_{k, n}\}_{n=0}^{N_{\tau}} $ where $ \tau^l_{k, n} = \frac{n}{N_{\tau}} (\tau_{k, \max}^l - \tau_{k, \min}^l) + \tau_{k, \min}^l $.
		\STATE Derive $ \{J^l_{k, n}\}_{n=0}^{N_{\tau}} $ with $ \{\tau^l_{k, n}\}_{n=0}^{N_{\tau}} $ by \eqref{equ:P_tau}.
		\STATE Derive $ n^* = \arg \min_n J^l_{k, n} $.
		\STATE $ \tau_{k, \min}^{l+1} \leftarrow \tau^l_{k, n} |_{n = \max(0, n^* - 1)} $.
		\STATE $ \tau_{k, \max}^{l+1} \leftarrow \tau^l_{k, n} |_{n = \min(N_{\tau}, n^* + 1)} $.
		\ENDFOR
		\STATE \emph{Output:} Estimated CSI uncertainty $ \hat{\tau}_k = \tau^{L_{\tau}}_{k, n}|_{n = n^*} $.
	\end{algorithmic}
	\label{alg:CUS}
\end{algorithm}

\subsection{Optimizations}

\subsubsection{Transmit RZF Precoding} 

The SR maximization problem with respect to regularization term $ \alpha $ is formulated as
\begin{equation}
\begin{split}\label{equ:P_R}
& \max_{\alpha} R_{\textup{p}}\\
& \, \textup{s.t.} \,\,\,\, \gamma_{\textup{p}, k} = g_{\gamma}(\boldsymbol{\mathcal{X}}_k; \boldsymbol{\Theta}_g)\\
& \,\,\,\,\,\,\,\,\,\,\,\, R_{\textup{p}} = \sum_{k=1}^K \log(1 + \gamma_{\textup{p}, k})
\end{split}
\end{equation}
where $ g_{\gamma} $ is the branch of function $ g $ which predicts SINR. According to conjecture and the analysis in \cite{6172680}, the learning-aided $ R_{\textup{p}} $ has a global maximum over $ \alpha \in \mathbb{R}^+ $ in the Case $ 4 $, but is non-convex in the Cases $ 1 $ to $ 3 $. Meanwhile, a closed-form derivation of stationary points is difficult for current learning tools such as tensorflow. Similarly, we resort to the searching method for regularization term optimization, and the proposed transmit RZF precoding optimization algorithm is given in Algorithm~\ref{alg:RTS}.

Finally, we replace the regularization term in $ \boldsymbol{\mathcal{X}}_k $ with the optimized $ \alpha^* $, and the corresponding numerical vector is also updated. The BS precodes downlink signal with $ \alpha^* $ in Fig.~\ref{fig:information_interaction}. The obtained updated $ \boldsymbol{\mathcal{X}}_k $ is the input vector for the next cascaded receive power scaling. 

\begin{algorithm}
	\caption{Transmit RZF precoding optimization algorithm.}
	\begin{algorithmic}[1]
		\STATE \emph{Input:} Estimated $ \{\hat{\tau}_k\}_{k=1}^K $, input vector set $ \{\boldsymbol{\mathcal{X}}_{k}\}_{k=1}^K $, maximum and minimum regularization terms $ \alpha_{\max}, \alpha_{\min} $, division number $ N_{\alpha} $, maximum iteration $ L_{\alpha} $.
		\STATE \emph{Initialization:} Update the input vector set $ \{\boldsymbol{\mathcal{X}}_{k}\}_{k=1}^K $ with $ \{\hat{\tau}_k\}_{k=1}^K $.
		\FOR{$ l = 1 $ to $ L_{\alpha} $}
		\STATE Derive $ \{\alpha^l_{n}\}_{n=0}^{N_{\alpha}} $ where $ \alpha^l_{n} = \frac{n}{N_{\alpha}} (\alpha_{\max}^l - \alpha_{\min}^l) + \alpha_{\min}^l $.
		\STATE Update the numerical vector $ \boldsymbol{N}_{k, n}^l $ with $ \alpha_{n}^l $ for all users.
		\STATE Update the input vector set $ \{\boldsymbol{\mathcal{X}}_{k, n}^l\}_{k=1}^K $ with $ \{\boldsymbol{N}_{k, n}^l\}_{k=1}^K $.
		\STATE Derive $ \{R^l_{\textup{p}, n}\}_{n=0}^{N_{\alpha}} $ with $ \{\boldsymbol{\mathcal{X}}_{k, n}^l\}_{k=1}^K $ by \eqref{equ:P_R}.
		\STATE Derive $ n^* = \arg \max_n R^l_{\textup{p}, n} $.
		\STATE $ \alpha_{\min}^{l+1} \leftarrow \alpha^l_{n} |_{n = \max(0, n^* - 1)} $.
		\STATE $ \alpha_{\max}^{l+1} \leftarrow \alpha^l_{n} |_{n = \min(N_{\alpha}, n^* + 1)} $.
		\ENDFOR
		\STATE \emph{Output:} Optimized regularization term $ \alpha^* = \alpha^{L_{\alpha}}_{n}|_{n = n^*} $.
	\end{algorithmic}
	\label{alg:RTS}
\end{algorithm}

\subsubsection{Receive Power Scaling}

The MSE minimization problem with respect to scaling factor set $ \{v_k\}_{k=1}^K $ can be decoupled among users. With the updated $ \boldsymbol{\mathcal{X}}_{k} $ as the initialized input, then the $ k $-th MSE minimization sub-problem is formulated as
\begin{equation}
\begin{split}\label{equ:P_MSE}
& \min_{v_k} \textup{MSE}_{\textup{p}, k}\\
& \, \textup{s.t.} \,\,\,\, \textup{MSE}_{\textup{p}, k} = g_{\textup{MSE}}(\boldsymbol{\mathcal{X}}; \boldsymbol{\Theta}_g)
\end{split}
\end{equation}
where $ g_{\textup{MSE}} $ is the branch of function $ g $ which predicts MSE. The optimized receive power scaling factor $ v_k^* $ is then transmitted to user $ k $ for detection. According to conjecture and the analysis in \cite{6172680}, the learning-aided $ \textup{MSE}_{\textup{p}, k} $ has a global minimum over $ v_k \in \mathbb{R}^+ $ in all the cases. Thus, we adopt PGD for receive scaling factor optimization. During each iteration, the power scaling factor $ v_k $ is updated with gradient $ \nabla_{v_k} \textup{MSE}_{\textup{p}, k} $, and then projected onto feasible region $ \mathbb{R}^+ $.

\begin{figure}[h!]
	\centering
	\includegraphics[width = 3.0in]{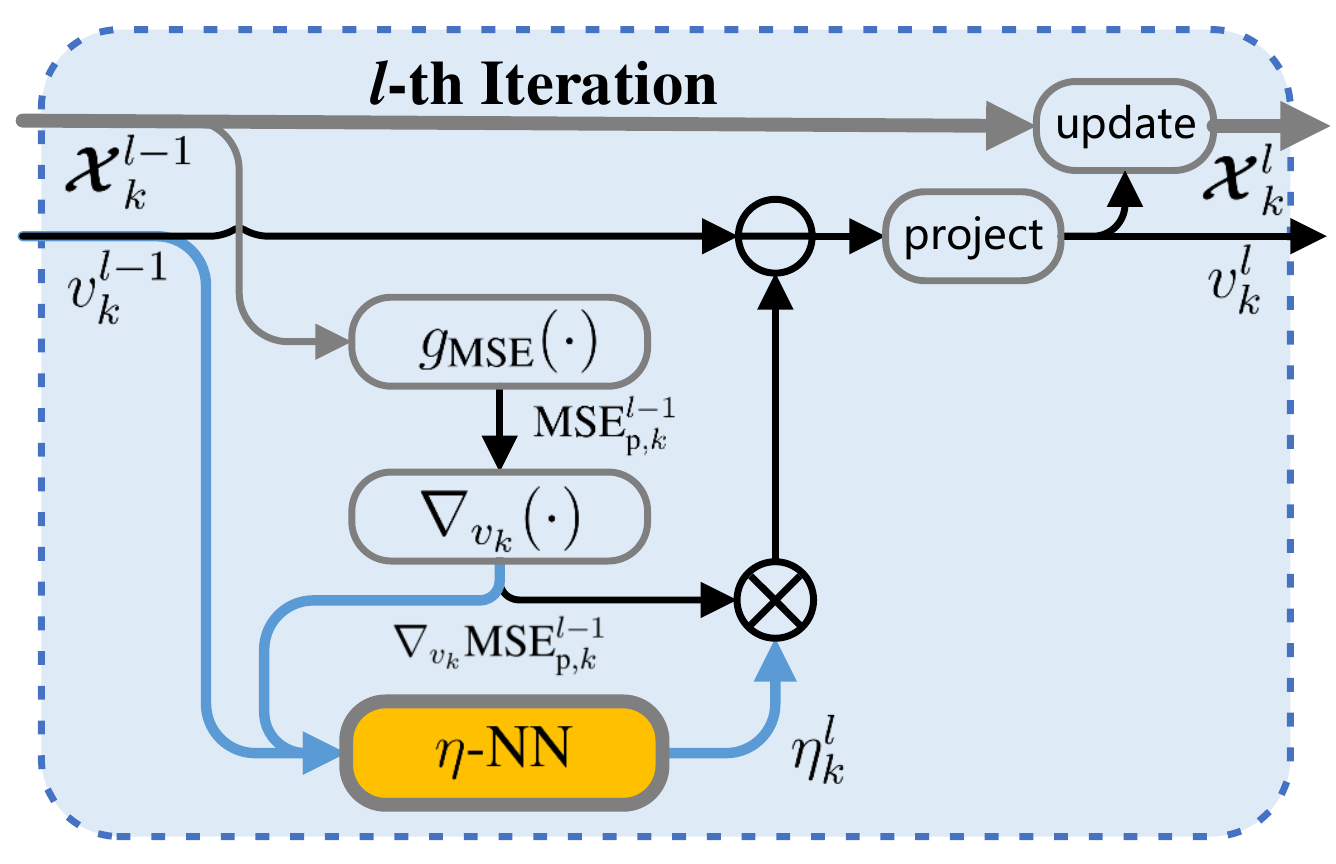}
	\caption{The $ l $-th iteration of deep unfolded PGD for receive power scaling optimization. The learnable part is labeled with orange.}
	\label{fig:deep_unfolded_PGD}
\end{figure}

In PGD, the update rate $ \eta $ is performed as a hyper-parameter, which lacks theoretical guidance and is usually designed by engineering experience. In practice, a small $ \eta $ stabilizes the convergence, but it leads to slow convergence speed; meanwhile, a large $ \eta $ can cause oscillation. To obtain a good trade-off between convergence speed and robustness, an adaptive update rate is preferred but is difficult to design. Thus, we propose a deep unfolded PGD which enables the adaptive update rate design to be data-driven. The adaption is realized by a NN with parameter set $ \boldsymbol{\Theta}_{\eta} $, which is called as $ \eta $-NN. The $ l $-th iteration of power scaling factor $ v_k^l $ is shown in Fig.~\ref{fig:deep_unfolded_PGD}. Specifically, $ \eta $-NN adaptively adjusts the update rate according to current $ v_k $ and $ \nabla_{v_k} \textup{MSE}_{\textup{p}, k} $. 
Thus, the input of $ \eta $-NN is $ [v_k, \nabla_{v_k} \textup{MSE}_{\textup{p}, k}]^{\textup{T}} $, and the corresponding output is $ \eta_k $. In a supervised learning manner, the training problem of deep unfolded PGD is formulated as
\begin{equation}
\begin{split}
& \min_{\boldsymbol{\Theta}_{\eta}} \mathbb{E}_{v_k^0 \sim \boldsymbol{\mathcal{V}}} \Big\{ \sum_{l=1}^L \textup{MSE}_{\textup{p}, k}^{l} \Big\}\\
& \, \textup{s.t.} \,\,\,\, \eta_k^{l} = h\Big(\big[v_k^{l-1}, \nabla_{v_k} \textup{MSE}_{\textup{p}, k}^{l-1} \big|_{v_k = v_k^{l-1}}\big]; \boldsymbol{\Theta}_{\eta}\Big)\\
& \,\,\,\,\,\,\,\,\,\,\,\, v_k^{l} = \max(v_k^{l-1} - \eta_k^{l} \nabla_{v_k} \textup{MSE}_{\textup{p}, k}^{l-1} \big|_{v_k = v_k^{l-1}}, 0)
\end{split}\label{equ:P_PGD_unfold}
\end{equation}
where $ h $ is the map of $ \eta $-NN, and the corresponding topology is illustrated in Table~\ref{tab:eta_net}. The activation function in the output layer bounds the update rate within range $ [10^{-3}, 10^{-1}] $.

\begin{table}[h!]
	\centering
	\setlength{\tabcolsep}{2.0mm}{
	\small
	\centering
	\caption{Topology of the $ \eta $-NN.} 
	\begin{tabular}{c|c} 
		\toprule
		Output layer & $ 10^{\max(\min(\cdot, -1), -3)} $, $ 1 $\\
		Hidden layer & ReLU, $ 8 $\\
		Input layer & linear, $ 2 $\\
		\bottomrule 
	\end{tabular}
	\label{tab:eta_net}}
\end{table}

During training, $ \Theta_{\eta} $ is iteratively updated by mini-batch gradient descent (MBGD) until convergence. Using the learned PP model $ g $, the training of $ h $ is endogenous and requires no external data. The proposed receive power scaling optimization algorithm is given in Algorithm~\ref{alg:PSO}. According to Appendix~\ref{sec:appendix_b}, the optimized power scaling factor $ u_k^* $ is computed with $ v_k^* $, and then transmitted from the BS to user $ k $ for detection in Fig.~\ref{fig:information_interaction}.

\begin{algorithm}
	\caption{Receive power scaling optimization algorithm (Deep unfolded PGD).}
	\begin{algorithmic}[1]
		\STATE \emph{Input:} Estimated $ \hat{\tau}_k $, input vector $ \boldsymbol{\mathcal{X}}_{k} $, maximum iteration $ L $.
		\STATE \emph{Initialization:} Update the input vector $ \boldsymbol{\mathcal{X}}_{k} $ with $ \hat{\tau}_k $; then update $ \boldsymbol{\mathcal{X}}_{k} $ with the optimized $ \alpha^* $ and the corresponding numerical vector $ \boldsymbol{N}_k $, and finally obtain $ \boldsymbol{\mathcal{X}}_{k}^0 $.
		\FOR{$ l = 1 $ to $ L $}
		\STATE Derive $ \textup{MSE}_{\textup{p}, k}^l $ by $ g_{\textup{MSE}} $ with $ \boldsymbol{\mathcal{X}}_{k}^{l-1} $.
		\STATE Compute gradient $ \nabla_{v_k} \textup{MSE}_{\textup{p}, k}^{l-1} \big|_{v_k = v_k^{l-1}} $.
		\STATE Derive $ \eta_k^l $ by $ \eta $-NN with input $ [v_k^l, \nabla_{v_k} \textup{MSE}_{\textup{p}, k}^l]^{\textup{T}} $.
		\STATE $ v_k^l \leftarrow v_k^{l-1} - \eta_k^{l} \nabla_{v_k} \textup{MSE}_{\textup{p}, k}^{l-1} \big|_{v_k = v_k^{l-1}} $.
		\STATE $ v_k^l = \max(v_k^l, 0) $.
		\STATE Update the input vector $ \boldsymbol{\mathcal{X}}_{k}^{l-1} $ with $ v_k^l $, and obtain $ \boldsymbol{\mathcal{X}}_{k}^l $.
		\ENDFOR
		\STATE \emph{Output:} Optimized $ v_k^* $.
	\end{algorithmic}
	\label{alg:PSO}
\end{algorithm}

\subsection{Computation and Storage Complexities}

The overall computation and storage complexities of learning-enabled models in the proposed dual-driven scheme are summarized in Table~\ref{tab:complexity}. The main computation complexity is caused by multiplication. The multiplication times in a single hidden-layer (HL) PP model, a double HL PP model, and an $ \eta $-NN respectively are up to hundreds, thousands, and tens. Additionally, the corresponding storage cost is at the same level. Thus, the computation and storage complexities of the proposed scheme is very small.

\begin{table}[h!]
	\centering
	\setlength{\tabcolsep}{1.5mm}{
		\small
		\centering
		\caption{Computation and storage complexities of learning-enabled parts in the dual-driven scheme.}
		\begin{tabular}{c|ccccc} 
			\toprule
			\multirow{2}{*}{NN model} & \multirow{2}{*}{multiplication} & \multirow{2}{*}{addition} & comparative & exponential & trainable\\
			& & & operator & operator & parameter\\
			\midrule
			PP (single HL) & $ 33 \overline{\boldsymbol{\mathcal{X}}}_k + 162 $ & $ \overline{\boldsymbol{\mathcal{X}}}_k + 70 $ & $ 32 $ & $ / $ & $ 34 \overline{\boldsymbol{\mathcal{X}}}_k + 228 $\\
			PP (double HL) & $ 33 \overline{\boldsymbol{\mathcal{X}}}_k + 1218 $ & $ \overline{\boldsymbol{\mathcal{X}}}_k + 134 $ & $ 64 $ & $ / $ & $ 34 \overline{\boldsymbol{\mathcal{X}}}_k + 1348 $\\
			$ \eta $-NN & $ 24 $ & $ 9 $ & $ 10 $ & $ 1 $ & $ 33 $\\
			\bottomrule
		\end{tabular}
		\label{tab:complexity}}
\end{table}

\section{Simulation Results}\label{sec:simulation}

\subsection{System Configurations}

The value range or set of BS antenna number $ M $, user number $ K $, CSI uncertainty $ \tau $ and maximal power $ P $ are given in Table~\ref{tab:sim}, $ \tau $ and $ P $ are uniformly distributed within the range. In Cases $ 1 $ to $ 3 $, the user power is randomly allocated. The precoded signal is transmitted by frame, and each observation is an average of $ 5000 $ frames. The element in the channel correlation matrix is modeled as 
\begin{equation}
\boldsymbol{\Theta}_{i,j} = \left\{
\begin{array}{lr}
r^{j-i}, & i \leq j,\\
\boldsymbol{\Theta}^*_{j,i}, & i > j,
\end{array}
\right.\nonumber
\end{equation}
where $ r $ is evenly distributed in the unit circle on the complex plane. Except for the underestimated MSE, we simulate the interference as some fixed constant and $ \rho = 8 \rho_{\textup{m}} $ is considered. The data and model-driven schemes (respectively labeled as `data-driven' and `model-driven') perform as baselines, and they have the same procedure in Fig.~\ref{fig:overall_framework} as the dual-driven scheme. Considering PP, CSI uncertainty estimation and subsequent optimizations, `data-driven' uses the learned PP module in Fig.~\ref{fig:predict_net}, without model-based results as auxiliary information; `model-driven' only uses the model-based part, without further refinement by the data-driven part. The simulation platform is presented as: Python 3.5, Tensorflow 1.14.0, CPU Intel i7-9700K and GPU Nvidia GTX-1070Ti.

\begin{table}[h!]
	\centering 
	\setlength{\tabcolsep}{2.0mm}{
	\small
	\centering
	\caption{Simulation Configurations} 
	\begin{tabular}{cc||cc} 
		\toprule
		Name & Value & Name & Value \\ 
		\midrule
		$ M $ set (Case $ 1 $) & $ \{2, 4\} $ & $ K $ set (Case $ 1 $) & $ \{2\} $\\
		$ M $ set (Cases $ 2 $ to $ 4 $) & $ \{2, 4, 8\} $ & $ K $ set (Cases $ 2 $ to $ 4 $) & $ \{2, 4\} $\\
		$ \tau $ range & $ [0.1, 0.4] $ & $ P $ range (dB) & $ [6, 20] $\\
		Modulation type & QPSK & Frame size & $ 256 $\\		
		\bottomrule
	\end{tabular}
	\label{tab:sim}}
\end{table}

\subsection{Fitting Error}

\begin{figure}[h!]
	\centering
	\includegraphics[width = 3.6in]{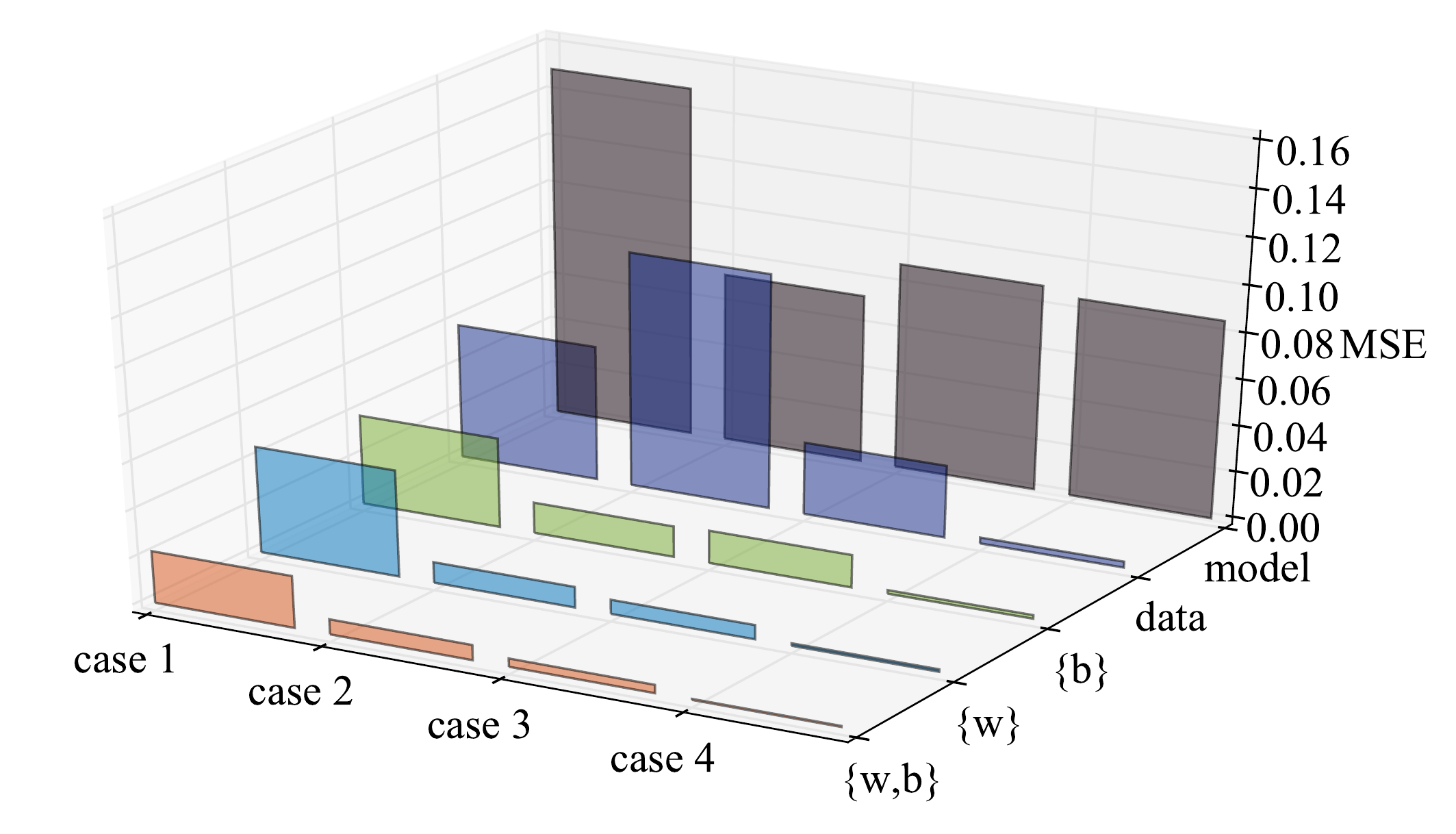}
	\caption{The fitting errors in interference free environments.}
	\label{fig:fitting_error}
\end{figure}

The core task of prediction models is to precisely predict the set of performance indicators. We use square fitting error as the measure. Firstly, considering perfectly observed MSE and no interference, the fitting error results are shown in Fig.~\ref{fig:fitting_error}. The notations $ \{w, b\} $, $ \{w\} $, $ \{b\} $ respectively are the models with weights and biases, only with weights, only with biases, and we name them as dual-driven models. Generally, the dual-driven methods have the best fitting performance on the four cases, compared to the model or data-driven methods. As the case is further reduced, the fitting error of dual-driven models decreases. The most significant fitting gain is achieved in Case $ 4 $, the MSEs of $ \{w, b\} $ and `model' respectively are $ 4.31 \times 10^{-4} $ and $ 8.56 \times 10^{-2} $.

Secondly, the fitting error results with underestimated MSE and interference are list in Table~\ref{tab:fitting}. Similarly, the models with $ \{w, b\} $ have the lowest MSEs (highlighted in bold fonts), and the model-driven methods have the highest MSEs (marked with color grey). The models trained with underestimated MSE and interference will be further tested in the next simulations.

\begin{table}[h!]
	\centering
	\setlength{\tabcolsep}{2.0mm}{$  $
	\small
	\centering
	\begin{threeparttable}
	\caption{The fitting errors in the scenario with underestimated MSE and interference.\tnote{1}}
	\begin{tabular}{c|ccc|c|c} 
		\toprule
		& \multicolumn{3}{c|}{dual-driven} & data- & model-\\ 
		& $ \{w, b\} $ (proposed) & $ \{w\} $ & $ \{b\} $ & driven & driven\\
		\midrule
		Case $ 1 $\tnote{2} & \bm{$ 2.25 \times 10^{-2} $} & $ 5.36 \times 10^{-2} $ & $ 5.01 \times 10^{-2} $ & $ 8.02 \times 10^{-2} $ & {\color[RGB]{130, 130, 130}$ 1.97 \times 10^{-1} $}\\
		Case $ 2 $ & \bm{$ 4.72 \times 10^{-2} $} & $ 6.57 \times 10^{-2} $ & $ 7.91 \times 10^{-2} $ & $ 1.28 \times 10^{-1} $ & {\color[RGB]{130, 130, 130}$ 1.81 \times 10^{-1} $}\\
		Case $ 3 $ & \bm{$ 4.57 \times 10^{-2} $} & $ 6.43 \times 10^{-2} $ & $ 6.93 \times 10^{-2} $ & $ 6.96 \times 10^{-2} $ & {\color[RGB]{130, 130, 130}$ 2.00 \times 10^{-1} $}\\
		Case $ 4 $\tnote{2} & \bm{$ 3.86 \times 10^{-4} $} & $ 1.03 \times 10^{-3} $ & $ 6.60 \times 10^{-4} $ & $ 8.81 \times 10^{-4} $ & {\color[RGB]{130, 130, 130}$ 7.32 \times 10^{-2} $}\\		
		\bottomrule
	\end{tabular}
	\label{tab:fitting}
    \begin{tablenotes}
	\footnotesize
	\item[1] Lower is better.
	\item[2] Cases $ 1 $ and $ 4 $ are fitted with double HL NNs, for better approximation performance.
	\end{tablenotes}
	\end{threeparttable}}
\end{table}

\subsection{CSI Uncertainty Estimation}

The division number and maximum iteration for CSI uncertainty estimation respectively are $ N_{\tau} = 10 $ and $ L_{\tau} = 2 $. The CSI uncertainty estimation results are given in Table~\ref{tab:estimation}, measured by MSE. In general, the dual-driven methods have achieved the lowest MSE, and among them our proposed model with $ \{w, b\} $ performs best (highlighted in bold fonts). Meanwhile, the dual-driven methods have lower $ \tau $ estimation error than the model and data-driven methods, and the model-driven methods have the worst estimation performance (marked with color grey).

\begin{table}[h!]
	\centering 
	\setlength{\tabcolsep}{2.0mm}{
	\small
	\centering
	\begin{threeparttable}
	\caption{The MSEs of CSI uncertainty estimation.\tnote{1}}
	\begin{tabular}{c|ccc|c|c} 
		\toprule
		& \multicolumn{3}{c|}{dual-driven} & data- & model-\\ 
		& $ \{w, b\} $ (proposed) & $ \{w\} $ & $ \{b\} $ & driven & driven\\ 
		\midrule
		case $ 1 $ & \bm{$ 3.43 \times 10^{-3} $} & $ 5.52 \times 10^{-3} $ & $ 4.60 \times 10^{-3} $ & $ 6.68 \times 10^{-3} $ & {\color[RGB]{130, 130, 130}$ 1.80 \times 10^{-2} $}\\
		case $ 2 $ & \bm{$ 1.99 \times 10^{-3} $} & $ 3.92 \times 10^{-3} $ & $ 2.99 \times 10^{-3} $ & $ 1.12 \times 10^{-2} $ & {\color[RGB]{130, 130, 130}$ 1.59 \times 10^{-2} $}\\
		case $ 3 $ & \bm{$ 6.16 \times 10^{-4} $} & $ 1.00 \times 10^{-3} $ & $ 1.01 \times 10^{-3} $ & $ 2.01 \times 10^{-3} $ & {\color[RGB]{130, 130, 130}$ 1.31 \times 10^{-2} $}\\
		case $ 4 $ & \bm{$ 1.98 \times 10^{-4} $} & $ 5.47 \times 10^{-4} $ & $ 6.92 \times 10^{-4} $ & $ 7.70 \times 10^{-4} $ & {\color[RGB]{130, 130, 130}$ 1.28 \times 10^{-2} $}\\	
		\bottomrule
	\end{tabular}
	\label{tab:estimation}
    \begin{tablenotes}
	\footnotesize
	\item[1] Lower is better.
	\end{tablenotes}
	\end{threeparttable}}
\end{table}

\subsection{Optimization Results}

The SR results optimized on the corresponding PP models are given in Table~\ref{tab:sumrate}. The division number and maximum iteration for precoding optimization respectively are $ N_{\alpha} = 10 $ and $ L_{\alpha} = 2 $. In each case, the highest ones are highlighted in bold fonts. The optimum derived by the iterative searching in Algorithm~\ref{alg:RTS} is labeled with $ (\cdot)^* $, where the PP model is replaced by the practical simulation scenario. In general, our proposed model with $ \{w, b\} $ in the dual-driven methods performs better or equal to the model-driven methods, and they have a smaller gap to the optimums. It can also be noticed that the SR of data-driven methods are worse than the model-driven methods, due to the inaccurate PP models.

After RZF precoding, the optimized detection results are given in Table~\ref{tab:mse}. The maximum iteration for power scaling is $ L_{v} = 5 $. Similarly, our proposed model with $ \{w, b\} $ in the dual-driven methods performs better or equal to the data or model-driven methods, and they have a smaller gap to the optimums. However, the detection performance of data or model-driven methods are not stable in all cases. The effectiveness of learnable model-driven PP-based optimizations are verified.

Additionally, we compare the performance with and without CSI uncertainty estimation both in Table~\ref{tab:sumrate} and Table~\ref{tab:mse}. The model-driven methods are adopted, and the constant $ 0.25 $ is used to replace $ \hat{\tau} $. We can observe that the performance is significantly improved with CSI uncertainty sensing and estimation.

\begin{table}[h!]
	\centering
	\setlength{\tabcolsep}{1.5mm}{
	\small
	\centering
	\begin{threeparttable}
	\caption{The SR with RZF precoding (bps/Hz).\tnote{1}}
	\begin{tabular}{c|cccc|ccc|c|cc|c} 
		\toprule
		& \multicolumn{4}{c|}{fixed $ \alpha $} & \multicolumn{3}{c|}{dual-driven} & data- & \multicolumn{2}{c|}{model-driven} & \multirow{2}{*}{optimal}\\
		& $ 0 $ & $ 0.01 $ & $ 0.1 $ & $ 1 $ & $ \{w, b\} $ & $ \{w\} $ & $ \{b\} $ & driven & $ \hat{\tau} $ & $ \hat{\tau} = 0.25 $ & \\ 
		\midrule
		Case $ 1 $ & $ 4.22 $ & $ 5.95 $ & $ 6.17 $ & $ 5.35 $ & \bm{$ 6.25 $} & $ 6.24 $ & $ 6.23 $ & $ 6.23 $ & $ 6.24 $ & $ 6.14 $ & $ 6.29^* $\\
		Case $ 2 $ & $ 3.73 $ & $ 6.58 $ & $ 6.96 $ & $ 5.66 $ & \bm{$ 7.10 $} & $ 7.08 $ & $ 7.09 $ & $ 6.92 $ & $ 7.00 $ & $ 6.76 $ & $ 7.14^* $\\
		Case $ 3 $ & $ 4.41 $ & $ 7.58 $ & $ 7.95 $ & $ 6.67 $ & \bm{$ 8.04 $} & $ 8.04 $ & $ 8.04 $ & $ 8.00 $ & $ 8.01 $ & $ 7.99 $ & $ 8.08^* $\\
		Case $ 4 $ & $ 5.58 $ & $ 5.90 $ & $ 6.48 $ & $ 5.37 $ & \bm{$ 6.57 $} & $ 6.56 $ & $ 6.56 $ & $ 6.50 $ & \bm{$ 6.57 $} & $ 6.14 $  & $ 6.61^* $\\	
		\bottomrule
	\end{tabular}
	\label{tab:sumrate}
    \begin{tablenotes}
	\footnotesize
	\item[1] Higher is better.
	\end{tablenotes}
	\end{threeparttable}}
\end{table}

\begin{table}[h!]
	\centering
	\setlength{\tabcolsep}{1.5mm}{
	\small
	\centering
	\begin{threeparttable}
	\caption{The MSEs in receive signal detection.\tnote{1}}
	\begin{tabular}{c|cccc|ccc|c|cc|c} 
		\toprule
		& \multicolumn{4}{c|}{fixed $ \alpha $} & \multicolumn{3}{c|}{dual-driven} & data- & \multicolumn{2}{c|}{model-driven} & \multirow{2}{*}{optimal}\\
		& $ 0 $ & $ 0.01 $ & $ 0.1 $ & $ 1 $ & $ \{w, b\} $ & $ \{w\} $ & $ \{b\} $ & driven & $ \hat{\tau} $ & $ \hat{\tau} = 0.25 $ & \\ 
		\midrule
		Case $ 1 $ & $ 0.99 $ & $ 0.42 $ & $ 0.27 $ & $ 0.42 $ & \bm{$ 0.24 $} & $ 0.25 $ & $ 0.25 $ & $ 0.26 $ & $ 0.26 $ & $ 0.28 $ & $ 0.22^* $\\
		Case $ 2 $ & $ 1.00 $ & $ 0.47 $ & $ 0.31 $ & $ 0.44 $ & \bm{$ 0.30 $} & $ 0.31 $ & \bm{$ 0.30 $} & $ 0.33 $ & $ 0.32 $ & $ 0.35 $ & $ 0.25^* $\\
		Case $ 3 $ & $ 1.00 $ & $ 0.43 $ & $ 0.29 $ & $ 0.41 $ & \bm{$ 0.27 $} & $ 0.29 $ & $ 0.29 $ & \bm{$ 0.27 $} & $ 0.28 $ & $ 0.30 $ & $ 0.22^* $\\
		Case $ 4 $ & $ 0.76 $ & $ 0.34 $ & $ 0.21 $ & $ 0.42 $ & \bm{$ 0.20 $} & $ 0.21 $ & $ 0.21 $ & $ 0.21 $ & \bm{$ 0.20 $} & $ 0.26 $ & $ 0.20^* $\\	
		\bottomrule
	\end{tabular}
	\label{tab:mse}
    \begin{tablenotes}
	\footnotesize
	\item[1] Lower is better.
	\end{tablenotes}
	\end{threeparttable}}
\end{table}

\subsection{Deep Unfolded PGD}

\begin{figure}[h!]
	\centering
	\includegraphics[width = 3.6in]{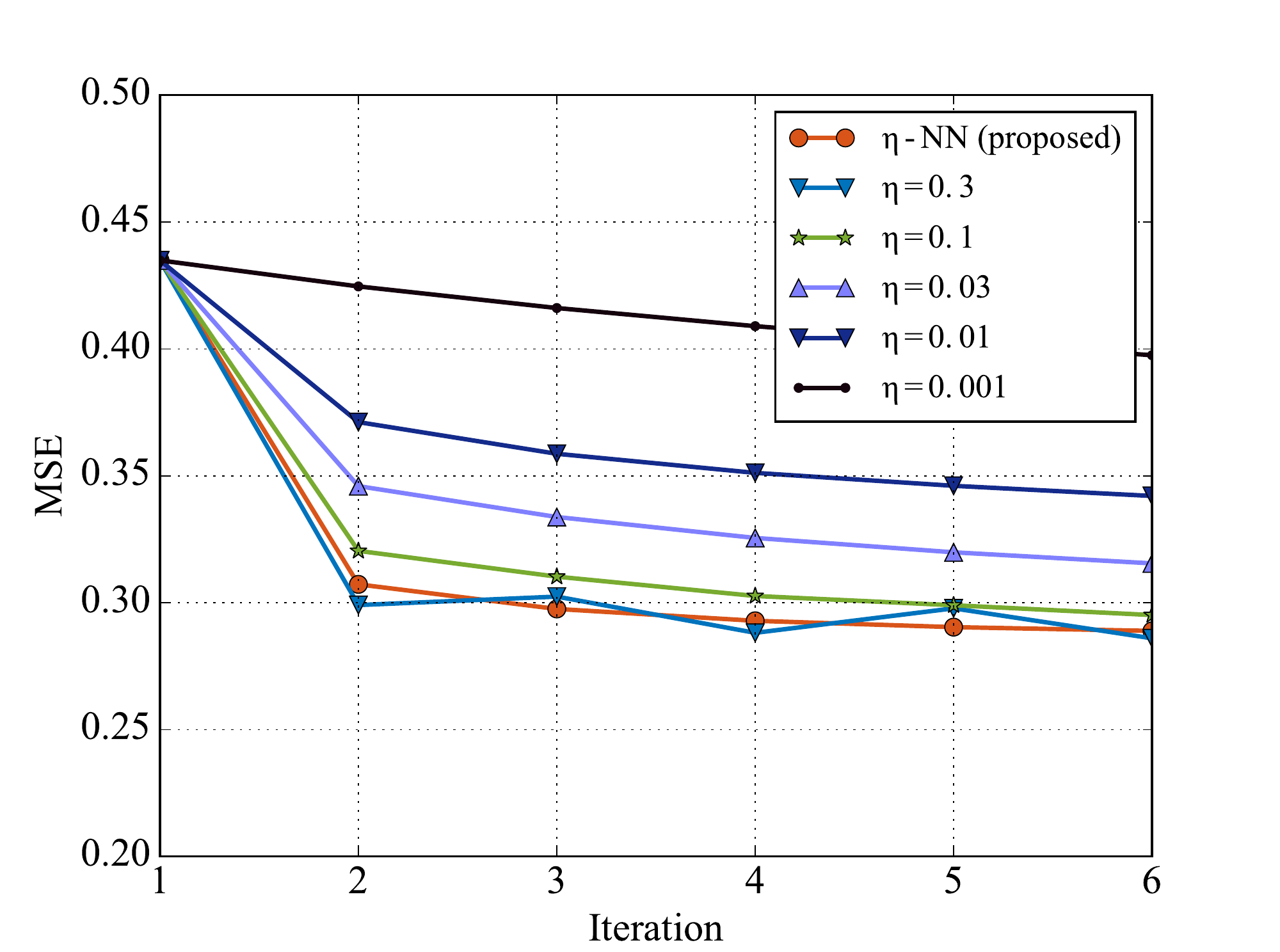}
	\caption{The virtual detection error versus iteration times.}
	\label{fig:eta}
\end{figure}

We take Case $ 3 $ as an instance, the virtual detection error (prediction by the model instead of practical measurement) versus iteration times $ L_{v} $ is shown in Fig.~\ref{fig:eta}. Generally, a small $ \eta $ has slow convergence speed; meanwhile, a large $ \eta = 0.3 $ causes oscillation. The proposed deep unfolded PGD with an adaptive $ \eta $, has a good trade-off between convergence speed and robustness. The simulation results verify the effectiveness of $ \eta $-NN.
     
\subsection{Imperfect SNR}

\begin{figure}[h!]
	\centering
	\subfigure[Sum rate]{
		\includegraphics[width=3.0in]{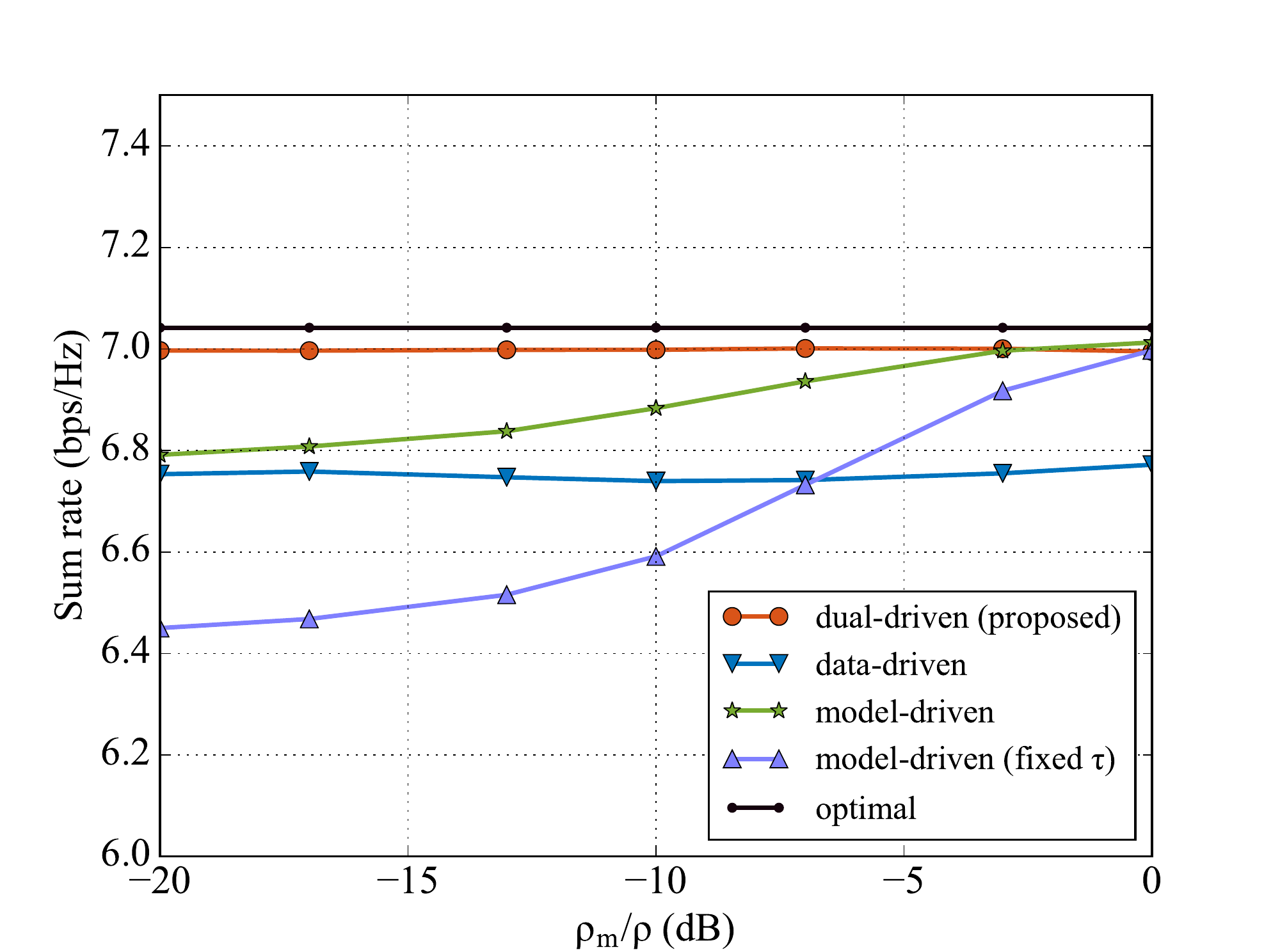}
		\label{fig:sum_rate_ratio}}
	\subfigure[MSE]{		
		\includegraphics[width=3.0in]{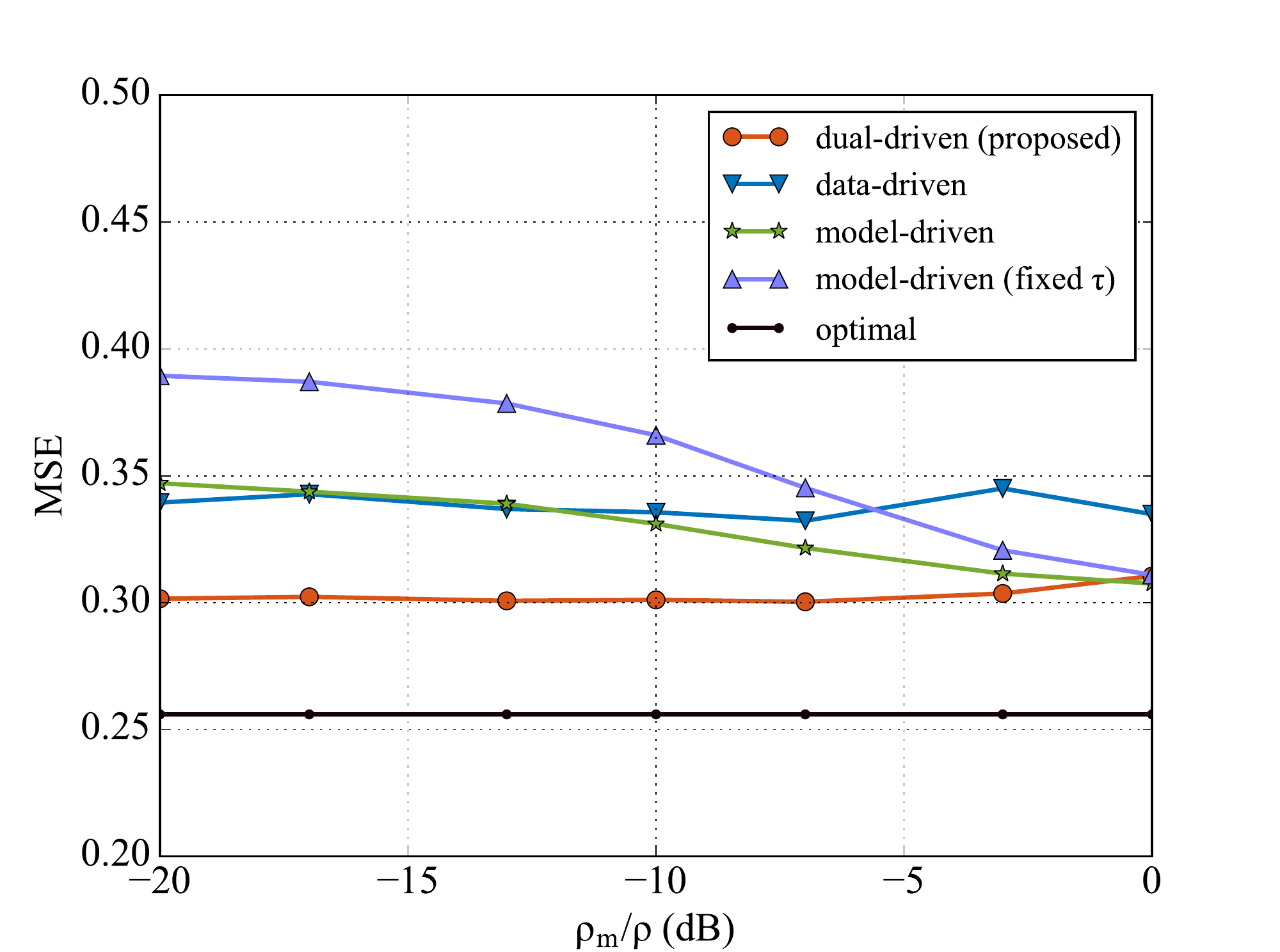}
		\label{fig:mse_ratio}}
	\caption{Optimizations versus the imperfect ratio.}
\end{figure}

We define imperfect ratio $ \rho_{\textup{m}}/\rho $ to reflect the degree of SNR imperfection, which is assumed to be in set $ \{-20, -17, -13, -10, -7, -3, 0\} \textup{dB} $. Clearly, a smaller $ \rho_{\textup{m}}/\rho $ means a more severe imperfection, and $ \rho_{\textup{m}}/\rho = 0 \textup{dB} $ indicates the observed SNR is perfect. As shown in Figs.~\ref{fig:sum_rate_ratio} and \ref{fig:mse_ratio}, the learning-enabled schemes are robust to the imperfect ratio. Meanwhile, as the $ \rho_{\textup{m}}/\rho $ becomes small, the SR of model-driven schemes is reduced and the MSE of detection grows. The proposed dual-driven scheme has achieved the best SR and MSE performance than the others. Since the distribution of practical SNR is unchanged, the SR and MSE of the optimal scheme are unchanged.

\subsection{Online Learning}

The $ \rho_{\textup{m}}/\rho $ follows uniform distribution in $ [-20, 0] \textup{dB} $, and is assumed to be i.i.d. across different time intervals and remains fixed within an time interval. At the beginning of each time interval, we re-train the PP model by supervised learning, and the learned PP performs in the rest of the time interval. The bottleneck of online learning is limited amount of dataset and high cost of computation in training. Thus, we evaluate the learned model with optimization performance, and also dataset size and training time cost.

\begin{figure}[h!]
	\centering
	\includegraphics[width = 3.6in]{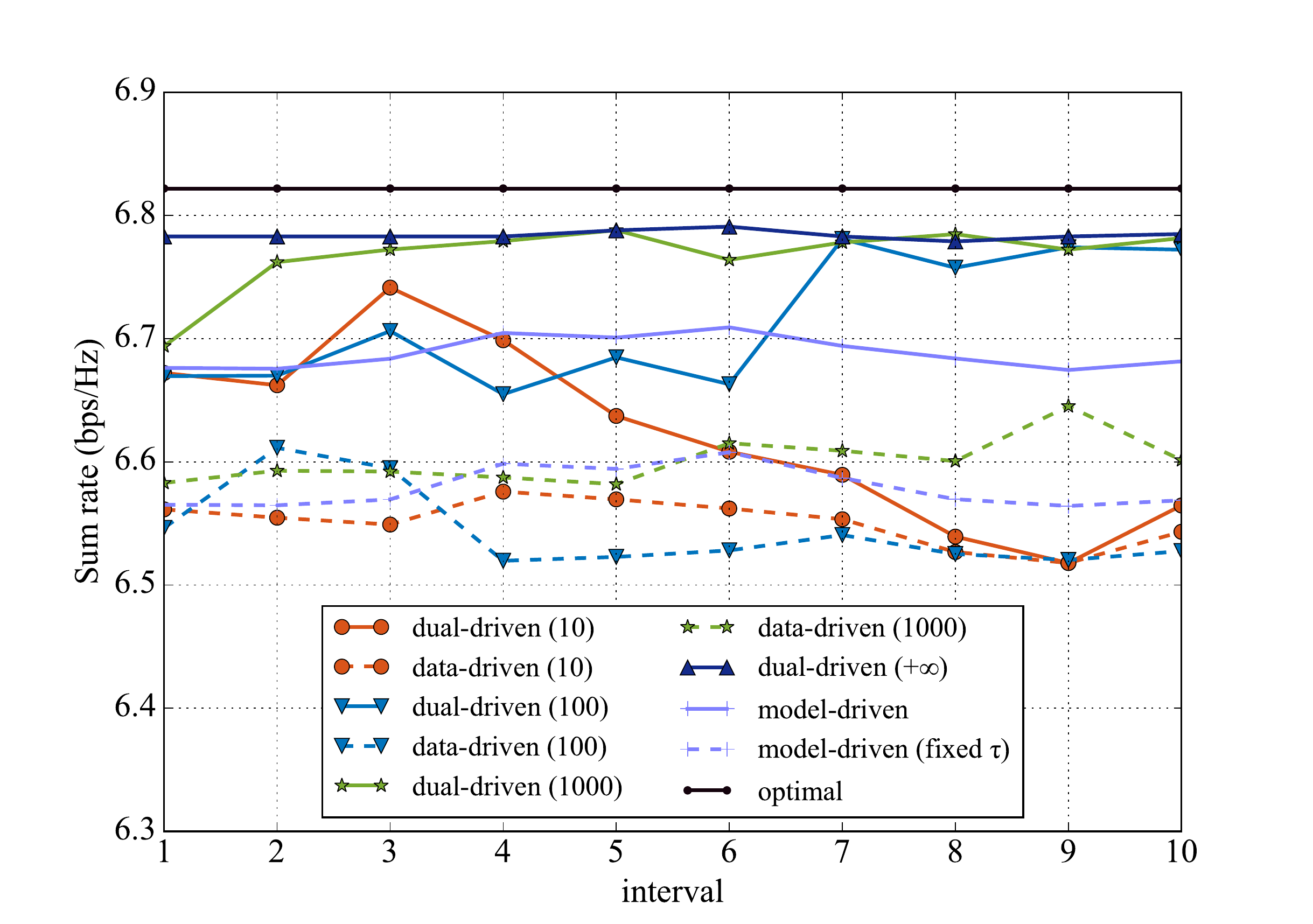}
	\caption{The sum rates versus time intervals.}
	\label{fig:sum_rate_interval}
\end{figure}

Considering initialization, the PP model is initialized with offline learned parameters in the first time interval, and with the latest learned parameters in the following time intervals. We take Case $ 2 $ with $ \{M = 4, K = 2\} $ as an example. As shown in Fig.~\ref{fig:sum_rate_interval}, the dual-driven and data-driven schemes respectively are plotted with solid and dashed lines. The digit in the brackets is the size of training dataset, and `$ +\infty $' denotes the size is sufficient large. The data-driven schemes cannot perform well with small dataset, while the dual-driven schemes significantly outperform the data-driven ones. When the size is up to $ 100 $, the dual-driven and model-driven schemes have similar SR performance. As the size grows to $ 1000 $, the dual-driven scheme outperforms the model-driven scheme, and is close to the one with very large dataset.

\section{Conclusions}\label{sec:conclusion}

The learnable model-driven framework for PP and subsequent optimization in imperfect MIMO system was introduced, then an application centered on transmit RZF precoding and receive signal detection with unknown channel uncertainty was given. Specifically, first we designed a light-weight NN for refined SR and detection MSE prediction based on coarse model-driven approximations. Using the learned PP model, second we adopted iterative searching method to estimate CSI uncertainty with measured indicators. Third, optimizations over the transmit RZF regularization term and cascaded receive power scaling factors were carried out. Besides, we designed a deep unfolded PGD algorithm for power scaling, to achieve faster convergence speed and better robustness than the vanilla PGD. The simulation results verified the effectiveness of proposed framework and methods, and showed that the designed predictor had more accurate performance predictions and CSI uncertainty estimation than the model-driven derivations. The optimized transmit RZF precoding had higher SR and receiver had lower detection MSE, respectively.

\begin{appendices}

\section{Deterministic Approximations of Performance}\label{sec:appendix_b}

Meanwhile, the closed-forms of SR and MSE are intractable to derive. Based on the high-dimensional random matrix theory, a deterministic equivalent of the SINR per user is derived in \cite{6172680}. Specifically, a close approximation independent of the the channel realization for every $ M $ which is (almost surely) exact as $ M \to \infty $ with finite $ \beta \triangleq M/K $. The instantaneous SINR $ \gamma_k $ is replaced by its large system approximation. The approximated variables are denoted with superscript $ (\cdot)^{\circ} $.

\subsection{General Case: $ \boldsymbol{\Theta}_k, \tau_k $}

In \cite{6172680}, the deterministic equivalents for the terms in \eqref{equ:gamma_k} and \eqref{equ:MSE_k} respectively are given as
\begin{align}
\boldsymbol{h}_k^{\textup{H}} \hat{\boldsymbol{W}} \hat{\boldsymbol{h}}_k & \stackrel{M \to \infty}{\longrightarrow} \sqrt{1-\tau_k^2} \frac{e_k}{1 + e_k}\label{equ:dq_1}\\
\boldsymbol{h}_k^{\textup{H}} \hat{\boldsymbol{W}} \hat{\boldsymbol{H}}_{[k]}^{\textup{H}} \boldsymbol{P}_{[k]} \hat{\boldsymbol{H}}_{[k]} \hat{\boldsymbol{W}} \boldsymbol{h}_k & \stackrel{M \to \infty}{\longrightarrow} \Upsilon_k^{\circ} \frac{(1 - \tau_k^2[1 - (1 + e_k)^2])}{(1 + e_k)^2}\label{equ:dq_2}\\
\xi^2 & \stackrel{M \to \infty}{\longrightarrow} \frac{P}{\Psi^{\circ}}\label{equ:dq_3}
\end{align}
Elements in $ \boldsymbol{e} = [e_1, \cdots, e_K]^{\textup{T}} $ form the unique positive solutions of $ e_i = \frac{1}{M} \textup{tr}(\boldsymbol{\Theta}_i \boldsymbol{T}) $ where $ \boldsymbol{T} = \bigg(\frac{1}{M} \sum_{j=1}^K \frac{\boldsymbol{\Theta}_j}{1+e_j} + \alpha \boldsymbol{I}_M \bigg)^{-1} $. Besides, approximations of $ \Psi^{\circ} $ and $ \Upsilon_k^{\circ} $ respectively are
\begin{align}
\Psi^{\circ} & \stackrel{M \to \infty}{\longrightarrow} \frac{1}{M} \sum_{j=1}^K \frac{p_j e'_j}{(1+e_j)^2} \quad \textup{and} \quad \Upsilon_k^{\circ} \stackrel{M \to \infty}{\longrightarrow} \frac{1}{M} \sum_{j=1, j \neq k}^K \frac{p_j e'_{j, k}}{(1+e_j)^2}\label{equ:upsilon_1}
\end{align}
with $ \boldsymbol{e}' = [e'_1, \cdots, e'_K]^{\textup{T}} $ and $ \boldsymbol{e}'_k = [e'_{1, k}, \cdots, e'_{K, k}]^{\textup{T}} $ given by
\begin{align}
\boldsymbol{e}' = (\boldsymbol{I}_K - \boldsymbol{J})^{-1} \boldsymbol{m} \quad \textup{and} \quad 
\boldsymbol{e}'_k = (\boldsymbol{I}_K - \boldsymbol{J})^{-1} \boldsymbol{m}_k\label{equ:e_1_4}
\end{align}
where the elements in $ \boldsymbol{J} $, $ \boldsymbol{m} $ and $ \boldsymbol{m}_k $ respectively take the form
\begin{align}
J_{i, k} & = \frac{1}{M^2 (1 + e_k)^2} \textup{tr}(\boldsymbol{\Theta}_i \boldsymbol{T} \boldsymbol{\Theta}_k \boldsymbol{T}) \quad \textup{and} \quad 
m_i = \frac{1}{M} \textup{tr}(\boldsymbol{\Theta}_i \boldsymbol{T}^2) \quad \textup{and} \quad 
m_{i, k} = \frac{1}{M} \textup{tr}(\boldsymbol{\Theta}_i \boldsymbol{T} \boldsymbol{\Theta}_k \boldsymbol{T})\label{equ:v_1_2}
\end{align}

Almost surely, the deterministic equivalents for SINR and MSE of user $ k $ respectively are derived as
\begin{align}
\gamma_k^{\circ} & \stackrel{M \to \infty}{\longrightarrow} \frac{p_k (1 - \tau_k^2) e_k^2}{\Upsilon_k^{\circ} (1 - \tau_k^2[1 - (1 + e_k)^2]) + \frac{\Psi^{\circ}}{\rho}(1 + e_k)^2}\label{equ:gamma_k_equ_1}\\
\textup{MSE}_k^{\circ} & \stackrel{M \to \infty}{\longrightarrow} \underbrace{\Big(u_k \sqrt{\frac{P}{\Psi^{\circ}}} \sqrt{p_k} \sqrt{1-\tau_k^2} \frac{e_k}{1 + e_k} - 1\Big)^2}_{\textup{residual signal power}} + \underbrace{u_k^2 \frac{P}{\Psi^{\circ}} \Upsilon_k^{\circ} \frac{(1 - \tau_k^2[1 - (1 + e_k)^2])}{(1 + e_k)^2}}_{\textup{interference power}} + \underbrace{u_k^2 \sigma^2}_{\textup{noise power}}\label{equ:mse_k_equ_1_1}
\end{align}
\eqref{equ:mse_k_equ_1_1} indicates that MSE is composed of three terms, (i) the residual signal power, (ii) the interference power, and (iii) the normalized noise power. We define $ v_k \triangleq \sqrt{\frac{P}{\Psi^{\circ}}} \sqrt{p_k} u_k $, then \eqref{equ:mse_k_equ_1_1} is further simplified as
\begin{equation}\label{equ:mse_k_equ_1}
\textup{MSE}_k^{\circ} \stackrel{M \to \infty}{\longrightarrow} \Big(v_k \sqrt{1-\tau_k^2} \frac{e_k}{1 + e_k} - 1\Big)^2 + v_k^2 \frac{\Upsilon_k^{\circ} (1 - \tau_k^2[1 - (1 + e_k)^2])}{p_k (1 + e_k)^2} + v_k^2 \frac{\Psi^{\circ}}{p_k \rho}
\end{equation}
The \eqref{equ:mse_k_equ_1} is quadratic over $ v_k $. The optimal $ v_k^* $ for minimizing \eqref{equ:mse_k_equ_1} is derived as
\begin{equation}\label{equ:v_1}
v_k^* = \frac{\sqrt{1-\tau_k^2} \frac{e_k}{1 + e_k}}{(1-\tau_k^2) \frac{e_k^2}{(1 + e_k)^2} + \frac{\Upsilon_k^{\circ} (1 - \tau_k^2[1 - (1 + e_k)^2])}{p_k (1 + e_k)^2} + \frac{\Psi^{\circ}}{p_k \rho}}
\end{equation}

\subsection{Reduced Channel Correlation Matrix: $ \boldsymbol{\Theta}_k = \boldsymbol{\Theta} $}

Channel correlation matrix is further reduced as $ \boldsymbol{\Theta}_k = \boldsymbol{\Theta}, \forall k $, then the element in $ \boldsymbol{e} $ is reduced to be the unique solution of
\begin{align}
e_i & = e = \frac{1}{M} \textup{tr}(\boldsymbol{\Theta} \boldsymbol{T}) \quad \textup{and} \quad \boldsymbol{T} = \bigg(\frac{\boldsymbol{\Theta}}{\beta (1+e)} + \alpha \boldsymbol{I}_M \bigg)^{-1}\label{equ:e_2_2}
\end{align}
We respectively define $ e_{12} $ and $ e_{22} $ as follows
\begin{align}
e_{12} & \triangleq \frac{1}{M (1+e)^2} \textup{tr}(\boldsymbol{\Theta} \boldsymbol{T}^2) \quad \textup{and} \quad e_{22} \triangleq \frac{1}{M (1+e)^2} \textup{tr}(\boldsymbol{\Theta} \boldsymbol{T} \boldsymbol{\Theta} \boldsymbol{T})\label{equ:e_22}
\end{align}
Then, the deterministic equivalents for $ \gamma_k $ and $ \textup{MSE}_k $ respectively are
\begin{align}
\gamma_k^{\circ} & \stackrel{M \to \infty}{\longrightarrow} \frac{\frac{p_k}{P / K} (1 - \tau_k^2) e [e_{22} + \alpha \beta (1+e)^2 e_{12}]}{(1 - \frac{p_k}{P}) (1 - \tau_k^2[1 - (1 + e)^2]) e_{22} + \frac{1}{\rho} (1 + e)^2 e_{12}}\label{equ:gamma_k_equ_2}\\
\textup{MSE}_k^{\circ} & \stackrel{M \to \infty}{\longrightarrow} \Big(v_k \sqrt{1-\tau_k^2} \frac{e}{1 + e} - 1\Big)^2 + v_k^2 \frac{(P - p_k) (1 - \tau_k^2[1 - (1 + e)^2]) e_{22}}{p_k K (1 + e)^2 (\beta - e_{22})} + v_k^2 \frac{\sigma^2 e_{12}}{p_k K (\beta - e_{22})}\label{equ:mse_k_equ_2}
\end{align}
The optimal $ v_k^* $ is derived as
\begin{equation}\label{equ:v_2}
v_k^* = \frac{\sqrt{1-\tau_k^2} \frac{e}{1 + e}}{(1-\tau_k^2) \frac{e^2}{(1 + e)^2} + \frac{(P - p_k) (1 - \tau_k^2[1 - (1 + e)^2]) e_{22}}{p_k K (1 + e)^2 (\beta - e_{22})} + \frac{\sigma^2 e_{12}}{p_k K (\beta - e_{22})}}
\end{equation}

\subsection{Uncorrelated Channel Matrix: $ \boldsymbol{\Theta}_k = \boldsymbol{I}_M $}

When $ \boldsymbol{\Theta}_k = \boldsymbol{I}_M $, $ e $ has a closed-form solution as follows 
\begin{equation}\label{equ:e_3_1}
e = \frac{\beta - 1 - \beta \alpha + \sqrt{(\beta - 1)^2 + 2 (1 + \beta) \alpha \beta + \alpha^2 \beta^2}}{2 \alpha \beta}
\end{equation}
The closed-form deterministic equivalents for $ \gamma_k $ and $ \textup{MSE}_k $ respectively are
\begin{align}
\gamma_k^{\circ} & \stackrel{M \to \infty}{\longrightarrow} \frac{\frac{p_k}{P / K} (1 - \tau_k^2) e [1 + \alpha \beta (1+e)^2]}{(1 - \frac{p_k}{P}) (1 - \tau_k^2 [1 - (1 + e)^2]) + \frac{1}{\rho} (1 + e)^2}\label{equ:gamma_k_equ_3}\\
\begin{split}
\textup{MSE}_k^{\circ} & \stackrel{M \to \infty}{\longrightarrow} \Big(v_k \sqrt{1-\tau_k^2} \frac{e}{1 + e} - 1\Big)^2 +\\
& v_k^2 \Big(\frac{(P - p_k) (1 - \tau_k^2[1 - (1 + e)^2])}{(1 + e)^2} + \sigma^2\Big) \cdot \frac{\beta}{p_k K ([\alpha \beta (1 + e) + 1]^2 - \beta)}
\end{split}\label{equ:mse_k_equ_3}
\end{align}
The optimal $ v_k^* $ is derived as
\begin{equation}\label{equ:v_3}
v_k^* = \frac{\sqrt{1-\tau_k^2} \frac{e}{1 + e}}{(1-\tau_k^2) \frac{e^2}{(1 + e)^2} + \Big(\frac{(P - p_k) (1 - \tau_k^2[1 - (1 + e)^2])}{(1 + e)^2} + \sigma^2\Big) \cdot \frac{\beta}{p_k K ([\alpha \beta (1 + e) + 1]^2 - \beta)}}
\end{equation}

\subsection{Uncorrelated $ \boldsymbol{\Theta} $ and Reduced $ \tau $: $ \boldsymbol{\Theta}_k = \boldsymbol{I}_M, \tau_k = \tau $}

The power allocation strategy maximizing the approximation of \eqref{equ:sumrate} is $ \boldsymbol{P}^{\ast, \circ} = \frac{P}{K} \boldsymbol{I}_M $. The allocated power for users are the same. Besides, $ e $, $ e' $ and $ e'' $ are already given in \eqref{equ:e_3_1} and \eqref{equ:e_3_2}. Therefore, the closed-form deterministic equivalents for $ \gamma_k $ and $ \textup{MSE}_k $ respectively are
\begin{align}
\gamma_k^{\circ} & = \gamma^{\circ} \stackrel{M \to \infty}{\longrightarrow} \frac{(1 - \tau^2) e [1 + \alpha \beta (1+e)^2]}{(1 - \frac{1}{K}) (1 - \tau^2 [1 - (1 + e)^2]) + \frac{1}{\rho} (1 + e)^2}\label{equ:gamma_k_equ_4}\\
\begin{split}
\textup{MSE}_k^{\circ} & = \textup{MSE}^{\circ} \stackrel{M \to \infty}{\longrightarrow} \Big(v \sqrt{1-\tau^2} \frac{e}{1 + e} - 1\Big)^2 +\\
& v^2 \Big(\frac{(1 - 1/K) (1 - \tau^2[1 - (1 + e)^2])}{(1 + e)^2} + \frac{1}{\rho}\Big) \cdot \frac{\beta}{([\alpha \beta (1 + e) + 1]^2 - \beta)}
\end{split}\label{equ:mse_k_equ_4}
\end{align}
where $ v_k = v $. When $ \boldsymbol{\Theta}_k = \boldsymbol{I}_M $ and $ \tau_k = \tau $, the SINR and MSE of different users are the same. Obviously, this conclusion also holds when $ \boldsymbol{\Theta}_k = \boldsymbol{I}_M $ is relaxed as $ \boldsymbol{\Theta}_k = \boldsymbol{\Theta} $. The optimal $ v^* $ is derived as
\begin{equation}\label{equ:v_4}
v^* = \frac{\sqrt{1-\tau^2} \frac{e}{1 + e}}{(1-\tau^2) \frac{e^2}{(1 + e)^2} + \Big(\frac{(1 - 1/K) (1 - \tau^2[1 - (1 + e)^2])}{(1 + e)^2} + \frac{1}{\rho}\Big) \cdot \frac{\beta}{([\alpha \beta (1 + e) + 1]^2 - \beta)}}
\end{equation}

\section{Concave Convex Analysis of \eqref{equ:P_tau}}\label{sec:appendix_c}

The partial differential equation (PDE) of $ J $ over $ \tau_k $ of user $ k $ is derived as
\begin{equation}
\begin{split}
\frac{\partial J_k}{\partial \tau_k} & = \frac{\partial J_{\gamma_k}}{\partial \tau_k} + \frac{\partial J_{\textup{MSE}_k}}{\partial \tau_k}
\end{split}\label{equ:J_div}
\end{equation}
Generally, the equation $ \frac{\partial J}{\partial \tau_k} = 0 $ is difficult to obtain, where $ J $ is an equal combination of $ J_{\gamma} $ and $ J_{\textup{MSE}} $. Considering an extended case where $ J_{\gamma} $ and $ J_{\textup{MSE}} $ can have inequal weights, a separate analysis is more appropriate. Thus, we make an analysis on the PDEs of $ J_{\gamma} $ and $ J_{\textup{MSE}} $, respectively. 
			
Firstly, the PDE of $ J_{\gamma_k} $ is derived as
\begin{align}
\frac{\partial J_{\gamma_k}}{\partial \tau_k} & = \frac{\partial (\gamma_{\textup{p}, k} - \gamma_{\textup{m}, k})^2}{\partial \tau_k} \approx \frac{\partial (\gamma_{k}^{\circ} - \gamma_{\textup{m}, k})^2}{\partial \tau_k}\nonumber\\
& = \bigg(\Big[\gamma_{\textup{m}, k} \big(\Upsilon_k^{\circ} + \frac{\Psi^{\circ}}{\rho}(1 + e_k)^2\big) - p_k e_k^2\Big] - \tau_k^2 \Big[\gamma_{\textup{m}, k} \Upsilon_k^{\circ} [1 - (1 + e_k)^2] - p_k e_k^2 \Big]\bigg) \cdot\nonumber\\
&\,\,\,\,\,\,\,\, \frac{2 \tau_k p_k e_k^2 (1 + e_k)^2 (\Upsilon_k^{\circ} + \frac{\Psi^{\circ}}{\rho})}{\big[\Upsilon_k^{\circ} (1 - \tau_k^2[1 - (1 + e_k)^2]) + \frac{\Psi^{\circ}}{\rho}(1 + e_k)^2\big]^3}\label{equ:J_gamma}
\end{align}
Obviously, $ \frac{\partial J_{\gamma_k}}{\partial \tau_k} = 0 $ has two solutions as
\begin{align}
\tau_{k, 1} & = \sqrt{\frac{\gamma_{\textup{m}, k} \big(\Upsilon_k^{\circ} + \frac{\Psi^{\circ}}{\rho}(1 + e_k)^2\big) - p_k e_k^2}{\gamma_{\textup{m}, k} \Upsilon_k^{\circ} [(1 + e_k)^2 - 1] + p_k e_k^2}}\label{equ:J_gamma_tau_1}\\
\tau_{k, 2} & = \sqrt{\frac{\Upsilon_k^{\circ} + \frac{\Psi^{\circ}}{\rho}(1 + e_k)^2}{\Upsilon_k^{\circ} [(1 + e_k)^2 - 1]}}\label{equ:J_gamma_tau_2}
\end{align}
where $ \tau_{k, 1} $ is the solution of $ \gamma_{k}^{\circ} = \gamma_{\textup{m}, k} $, and is also the global minimum of \eqref{equ:J_gamma} due to non-negativity of \eqref{equ:J_gamma}; but $ \tau_{k, 2} $ is a different zero point, and only exists when $ \frac{\Upsilon_k^{\circ} + \frac{\Psi^{\circ}}{\rho}(1 + e_k)^2}{\Upsilon_k^{\circ} [(1 + e_k)^2 - 1]} \leq 1 $. Thus, $ J_{\gamma} $ is non-convex over $ \tau $.

Secondly, the PDE of $ J_{\textup{MSE}_k} $ is derived as
\begin{align}
\frac{J_{\textup{MSE}_k}}{\partial \tau_k} & = \frac{\partial (\textup{MSE}_{\textup{p}, k} - \textup{MSE}_{\textup{m}, k})^2}{\partial \tau_k} \approx \frac{\partial (\textup{MSE}_{k}^{\circ} - \textup{MSE}_{\textup{m}, k})^2}{\partial \tau_k}\nonumber\\
& = \bigg[\Big(v_k \sqrt{1-\tau_k^2} \frac{e_k}{1 + e_k} - 1\Big)^2 + v_k^2 \frac{\Upsilon_k^{\circ} (1 - \tau_k^2[1 - (1 + e_k)^2])}{p_k (1 + e_k)^2} + v_k^2 \frac{\Psi^{\circ}}{p_k \rho} - \textup{MSE}_{\textup{m}, k}\bigg] \cdot\nonumber\\
&\,\,\,\,\,\,\,\, \tau_k \bigg[\frac{v_k e_k}{1 + e_k} \frac{1}{\sqrt{1 - \tau_k^2}} - \frac{v_k^2}{(1 + e_k)^2} \Big(e_k^2 + \frac{2 \Upsilon_k^{\circ}}{p_k} [1 - (1 + e_k)^2]\Big)\bigg]\label{equ:J_mse}
\end{align}
Obviously, $ \frac{\partial J_{\textup{MSE}_k}}{\partial \tau_k} = 0 $ has possible two solutions as
\begin{align}
\tau_{k, 1} & = \sqrt{1 - (1+e_k)^2 \bigg(\frac{\frac{e_k}{v_k} + \sqrt{\frac{e_k^2}{v_k^2} - \Big(e_k^2 + \frac{\Upsilon_k^{\circ}}{p_k} [1 - (1+e_k)^2]\Big) \Big(\frac{\Upsilon_k^{\circ} + \frac{\Psi^{\circ}}{\rho}}{p_k} + \frac{1 - \textup{MSE}_{\textup{m}, k}}{v_k^2}\Big)}}{e_k^2 + \frac{\Upsilon_k^{\circ}}{p_k} [1 - (1+e_k)^2]}\bigg)^2}\label{equ:J_mse_tau_1}\\
\tau_{k, 2} & = \sqrt{1 - \bigg(\frac{e_k (1 + e_k)}{v_k \big(e_k^2 + \frac{2 \Upsilon_k^{\circ}}{p_k} [1 - (1 + e_k)^2]\big)}\bigg)^2}\label{equ:J_mse_tau_2}
\end{align}
where $ \tau_{k, 1} $ is the solution of $ \textup{MSE}_{k}^{\circ} = \textup{MSE}_{\textup{m}, k} $, and is also the global minimum of \eqref{equ:J_mse} due to non-negativity of \eqref{equ:J_mse}. The solution $ \tau_{k, 2} $ only exists when $ 0 \leq \frac{e_k (1 + e_k)}{v_k \big(e_k^2 + \frac{2 \Upsilon_k^{\circ}}{p_k} [1 - (1 + e_k)^2]\big)} \leq 1 $. Therefore, $ J_{\textup{MSE}} $ is also non-convex over $ \tau $.

\end{appendices}

\bibliographystyle{IEEEtran}
\bibliography{References}

\end{document}